\documentclass[notitlepage,article,onecolumn,pre,aps,floatfix]{revtex4-1}
\usepackage{amsmath}
\usepackage{graphicx}
\usepackage{epstopdf}
\usepackage[usenames]{color}
\usepackage{xr}
\usepackage{placeins}
\usepackage[breaklinks=true,
            pdfborder={0 0 1},
            colorlinks=true,
            linkcolor=black,
            citecolor=blue,
            urlcolor=blue]{hyperref}
\usepackage[normalem]{ulem}
\usepackage{rotating}
\usepackage{verbatim}
\usepackage{bibentry}
\usepackage{xspace}
\usepackage[margin=0.5in]{geometry}
\usepackage{bm}
\usepackage{amssymb}
\usepackage{soul}

\pdfoutput=1

\usepackage[dvipsnames]{xcolor}

\usepackage{fancyhdr}
\begin{document}

\title{Steady-state distributions of ideal active Brownian particles under confinement and forcing}



\author{Caleb G. Wagner}
\author{Michael F. Hagan}
\email{hagan@brandeis.edu}
\author{Aparna Baskaran}
\email{aparna@brandeis.edu}

\affiliation{Martin Fisher School of Physics, Brandeis University,
  Waltham, Massachusetts 02453, USA.}

\begin{abstract}
We develop a formally exact technique for obtaining steady-state distributions of non-interacting active Brownian particles in a variety of systems. Our technique draws on results from the theory of two-way diffusion equations to solve the steady-state Smoluchowski equation for the 1-particle distribution function. The methods are employed to study in detail three scenarios: 1) confinement in a channel, 2) a constant flux steady state, and 3) sedimentation in a uniform external field.  In each scenario, known behaviors are reproduced and precisely quantified, and new results are presented. In particular,  in the constant flux state we derive an effective diffusivity which interpolates between the ballistic behavior of particle trajectories at short distances and their diffusive behavior at large distances. We also calculate the sedimentation profile of active Brownian particles near a wall, which complements earlier studies on the part far from the wall. Our techniques easily generalize to other active models, including systems whose activity is modeled in terms of Gaussian colored noise.
\end{abstract}

\maketitle

\section{Introduction}

Active matter describes a class of systems which are maintained far from equilibrium by driving forces acting on the constituent particles.
Experimental realizations of active matter are diverse and span several scales. Examples include the cell cytoskeleton \cite{Schaller2011}, bacterial suspensions \cite{Dombrowski2004,Kaiser2014}, synthetically prepared self-propelled colloids \cite{Palacci2010,Bricard2013,Narayan2006a,Narayan2007}, and flocking animals \cite{Attanasi2014b,Attanasi2014c,Holldobler1994}. Although in some cases active systems approach nonequilibrium steady states which can be described using concepts from equilibrium statistical mechanics \cite{Solon2015,Redner2016,Takatori2014,Takatori2015}, such an approach is not always possible \cite{Solon2015b}. A key reason for this limitation is that in many systems active motion remains correlated over length scales comparable to system size. Because of these correlations, the behavior of active systems under confinement or in the presence of obstacles shows a clear dependence on the details of the particle-wall interactions as well as wall geometry. Previous studies have revealed a number of striking behaviors that emerge in such systems, including spontaneous flow induced by asymmetric obstacles \cite{Tailleur2009,Wan2008,Angelani2009,Gosh2013}, accumulation of particles at walls \cite{Fily2014,Lee2013,Elgeti2013,Elgeti2009,Fily2015,Wensink2008}, and long-range depletion-induced forces \cite{Harder2014,Ni2015,Angelani2011}.




The persistent motion necessary to produce such novel phenomena is present even in a non-interacting model. Therefore, the associated (closed) equation for the 1-particle distribution function is by itself expected to capture many aspects of the rich behavior observed in such systems. These equations have a simple form. For instance, for active Brownian particles (ABPs) in 2d the steady-state distribution $f(\boldsymbol{r}, \theta)$ solves the (non-dimensional) equation
\begin{equation}
\hat{\boldsymbol{\nu}} \cdot \boldsymbol{\nabla} f(\boldsymbol{r}, \theta)   =\frac{\partial ^{2}f(\boldsymbol{r}, \theta)}{\partial \theta ^{2}},
\label{eq:1-FP-2d-nondimensional}
\end{equation}
where $\hat{\boldsymbol{\nu}} = (\cos \theta, \sin \theta)$ parametrizes particle orientations. Perhaps surprisingly, despite the simple form of this equation, its full solution appears difficult even in the simplest cases, and there is a need for a method of solution which is both formally exact as well as numerically accessible. Towards this end, Lee \cite{Lee2013} has considered ABPs confined in a channel (in which case Eq. (\ref{eq:1-FP-2d-nondimensional}) reduces to an effective 1d description) and has proposed an expansion in separable solutions. However, Lee leaves open a formal justification of the expansion, as well as the nontrivial problem of determining the expansion coefficients. Our goal here is to address both of these issues, and in doing so establish a rigorous yet numerically accessible technique for solving a broad class of equations of the form (\ref{eq:1-FP-2d-nondimensional}).  In particular, the technique we develop can be readily applied to systems whose spatial symmetry reduces the problem to an effective 1d description.


We demonstrate our techniques on two-dimensional ABPs in three cases. \emph{1) Confinement in a channel:} We precisely quantify known effects of confinement on active particles, including induced orientational order and large spatial gradients near the boundary. Moreover, we prove simple scaling relations (as a function of channel width) for both the bulk density and the fraction of particles adsorbed on the boundaries.   \emph{2) Constant flux steady state:} By considering a constant flux steady state between two reservoirs of fixed density, we demonstrate that the expansion in separable solutions in fact does not span the solution space --- a non-separable solution that is linear in the spatial variable is necessary to account for the nonzero particle flux. Using this solution, we derive an effective diffusivity for ABPs and discuss the signatures of activity. \emph{3) Sedimentation:} We calculate the full density profile for sedimenting ABPs in a uniform external field. Parametrizing in terms of a single spatial variable $x$, boundary data (such as determined by a hard wall) are imposed at the origin, and for $x > 0$ the particles are subject to a uniform force in the $-x$ direction. While the part of the distribution far from the wall can easily be obtained by assuming that the spatial and angular dependence in $f(x, \theta)$ factorizes \cite{Solon2015c}, calculation of the  part near the wall requires a more detailed analysis, which we present here.

Although we focus on ABPs, the techniques developed readily generalize to other active models as well, such as the more recently studied active-Ornstein-Uhlenbeck particles (AOUPs) \cite{Fodor2016} -- particles whose activity is modeled as Gaussian colored noise. In fact, we stress that our approach applies much more generally to a class of PDEs known in the mathematical literature as two-way diffusion equations \cite{Kruskal1980,Beals1981}. While it is known that the solutions to such equations can be rigorously expressed as an expansion in separable solutions \cite{Beals1981}, the explicit construction of the expansion remains difficult, largely due to the challenge of choosing expansion coefficients that satisfy boundary conditions. As part of our technique, we develop an iterative procedure which represents a possible approach to this issue on the finite domain. Numerical evidence from the ABP model as well as plausibility arguments indicate that this procedure does indeed converge to the exact solution.



\section{Statement of the Problem}
Non-interacting ABPs in two dimensions are parametrized by their position $\boldsymbol{r}$ and orientation $\theta$, which obey the overdamped Langevin equations:
\begin{align}
  \dot{\boldsymbol{r}} &= v_0 \hat{\boldsymbol{\nu}} + \sqrt{2 D} \boldsymbol{\eta}^\text{T}
  \label{eq:2-abp-spatial}
  \\
  \dot{\theta} &= \sqrt{2 D_r} \eta^\text{R}.
  \label{eq:2-abp-rotational}
\end{align}
Here $v_0$ is the magnitude of the self-propulsion velocity,
$\hat{\boldsymbol{\nu}} = (\cos \theta, \sin \theta)$, and $D$ and $D_r$ are the translational and rotational diffusion coefficients.  The
$\eta$ variables introduce Gaussian noise, with $\langle
\eta_i(t)\rangle = 0$ and $\langle \eta_i(t) \eta_j(t')\rangle =
\delta_{ij} \delta(t - t')$. To extract the physical signatures of activity in the simplest way possible, we ignore translational diffusion, setting $D = 0$. With this simplification, the only intrinsic length scale is the persistence length $\ell_p \equiv v_0 / D_r$; the ABPs can be viewed qualitatively as persistent random walkers.

To study the properties of this model in steady state, we work in terms of the steady-state Smoluchowski equation for the one-particle distribution function $f(\boldsymbol{r}, \theta)$,
\begin{equation}
\ell_p \hat{\boldsymbol{\nu}} \cdot \boldsymbol{\nabla} f(\boldsymbol{r}, \theta)   =\frac{\partial ^{2}f(\boldsymbol{r}, \theta)}{\partial \theta ^{2}},
\label{FP-2d-dimensional}
\end{equation}
which can be derived from Eqs. (\ref{eq:2-abp-spatial}) and (\ref{eq:2-abp-rotational}) using standard techniques \cite{Risken1989}. Restricting our focus to (quasi)-1d systems in rectangular coordinates, we may parametrize this equation in terms of the single spatial variable $x$:
\begin{equation}
\ell_p \cos \theta \frac{\partial f(x, \theta)}{\partial x}=\frac{\partial ^{2}f(x, \theta)}{\partial \theta ^{2}}.
\label{FP-1d-dimensional}
\end{equation}
Physically, this represents a continuity equation for ABPs on some volume $x \in [0, L]$. We therefore expect one set of well-posed boundary conditions to specify the particle fluxes incoming to this volume. In terms of the distribution function $f(x, \theta)$, this amounts to fixing
\begin{eqnarray}
f(0,\theta ) &=&v_{+}(\theta ),\text{ \ \ \ \ \ where } \cos(\theta) >0 \label{general-boundary-condition1} \\
f(L,\theta ) &=&v_{-}(\theta ),\text{ \ \ \ \ \ where } \cos(\theta) <0
\label{general-boundary-condition2}
\end{eqnarray}
for some functions $v_{\pm}$.

The system given by Eqs. (\ref{FP-1d-dimensional}), (\ref{general-boundary-condition1}), and (\ref{general-boundary-condition2}) in fact belongs to a broad class of problems known in the mathematical literature as two-way diffusion equations \cite{Kruskal1980}. The ``two-way" part of this name refers to the fact that particles drift to the right when $\cos \theta > 0$ and to the left when $\cos \theta < 0$, resulting in behavior very different from what is observed for typical second order diffusion operators. Previous studies have approached such equations using spectral techniques based on a separation of variables ansatz  (see, e.g. Refs. \cite{Freiling2003, Beals1981, Kruskal1980,Bethe1938}), which is the route we follow here as well.

\section{Solution by Separation of Variables}

\begin{figure}
  \includegraphics[width=0.4\linewidth,height=0.28\linewidth]{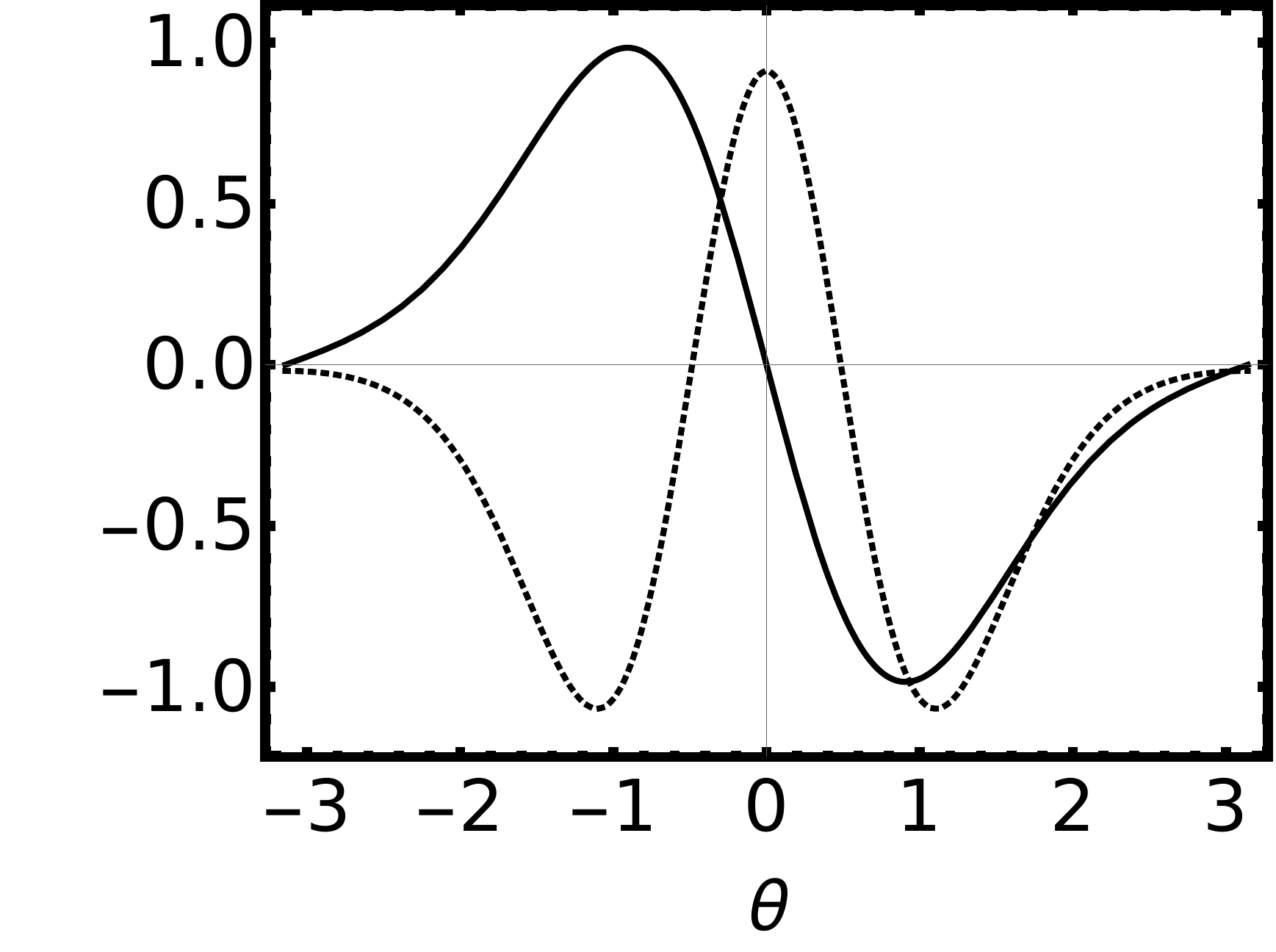}
  \includegraphics[width=0.4\linewidth,height=0.28\linewidth]{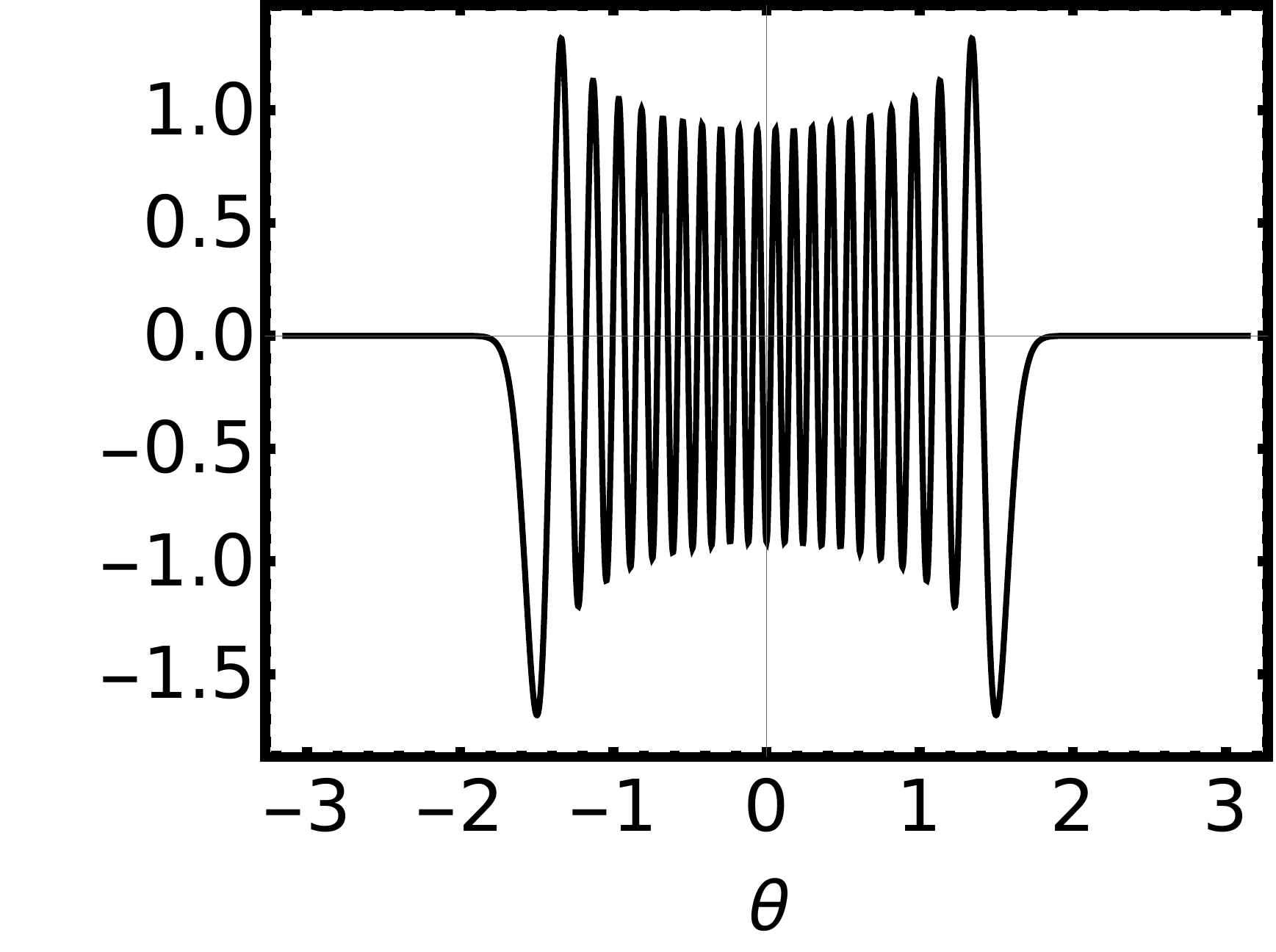}
  \caption{Plots of solutions to Eq. (\ref{eq:3-theta-eigenvalue-eqtn}). Left: $\Theta_{1}$ (solid black) and $\Theta_{2}$ (dashed black). Right: The $20^{th}$ even eigenfunction. The eigenfunctions are normalized such that $\int \Theta_k^2 \cos \theta d\theta = \text{sgn}(k)$ and $\Theta_{-k}(\theta) = \Theta_{k}(\theta + \pi)$.}
  \label{fig:eigenfunctions}
\end{figure}

To ease notation, in the present section we work in units where $\ell_p = 1$. Proceeding by separation of variables, we find that the separable solutions $\Gamma(x) \Theta(\theta)$ to Eq. (\ref{FP-1d-dimensional}) obey the ODEs
\begin{eqnarray}
\frac{d\Gamma }{dx} &=&\lambda\Gamma  \\
\frac{d^{2}\Theta }{d\theta ^{2}} &=&\lambda (\cos \theta) \Theta, \hspace{8 mm} \text{with} \hspace{2 mm} \Theta(\theta) = \Theta(\theta + 2\pi).
\label{eq:3-theta-eigenvalue-eqtn}
\end{eqnarray}
Because the weight function $\cos \theta$ is even, the angular eigenfunctions may be chosen to have definite parity.
The separable solution corresponding to $\lambda = 0$ is just a constant, which we call $\alpha$. For $\lambda \neq 0$, the $\Gamma(x)$ are simple exponentials, whereas the angular eigenfunctions $\Theta(\theta)$ are related to Mathieu functions. The latter may be efficiently constructed by separately expressing the odd and even eigenfunctions in terms of a $2\pi$ periodic Fourier expansion (see Appendix A). This construction automatically solves for the nonzero eigenvalues as well, which are real, discrete, and antisymmetric about $0$. Indexing each eigenvalue and its corresponding eigenfunction by $k$, we arrange the eigenvalues in descending order with $k$, adopting the convention that $\lambda_k$ is negative if $k$ is positive, and positive if $k$ is negative. Thus,
\begin{align}
\lambda_k &= -\lambda_{-k} \\
\cdots < \lambda_{2} < \lambda_{1} &< \lambda_{-1} < \lambda_{-2} < \cdots
\end{align}
A few sample eigenfunctions are shown in Fig. \ref{fig:eigenfunctions}, normalized such that $\int \Theta_k^2 \cos \theta d\theta = \text{sgn}(k)$ and $\Theta_{-k}(\theta) = \Theta_{k}(\theta + \pi)$. We note that the eigenfunctions with $k > 0$ are oscillatory where $\cos \theta > 0$ and exponentially decaying where $\cos \theta < 0$, with the reverse holding for eigenfunctions with $k < 0$.

While we would like to express the general solution of Eq. (\ref{FP-1d-dimensional}) as a sum of separable solutions, these do not span the solution space. There exists a \emph{diffusion solution} $\beta(x - \cos \theta)$ which cannot be expressed as a sum of separable solutions, and in fact is related to a double degeneracy in the $0$ eigenvalue \cite{Kruskal1980}. The physical significance of the diffusion solution can be inferred with the aid of the orthogonality property
\begin{equation}
 \int_{-\pi}^{\pi} \Theta_k \cos \theta d\theta = 0,
\end{equation}
 obtained by integrating both sides of Eq. (\ref{eq:3-theta-eigenvalue-eqtn}). This property implies that if we expand $f(x, \theta)$ as a sum of separable solutions plus the diffusion solution, and calculate the flux $\int v_0 \cos \theta f(x, \theta) d\theta$, we find that the flux is directly proportional to $\beta$ and independent of $x$. Therefore, the diffusion solution is required in order to correctly describe a constant flux steady state.


From the previous discussion, we may now write the general solution to Eq. (\ref{FP-2d-dimensional}) in the form
\begin{equation}
f(x, \theta) = \alpha + \beta (x - \cos \theta) + \sum_{k>0}a_{k}e^{\lambda _{k}x}\Theta_k+\sum_{k<0}a_{k}e^{\lambda _{k}\left( x-L\right) }\Theta_k.
\label{eq:3-two-way-sol-expansion}
\end{equation}
Note that without loss of generality we have split the sum over separable solutions into a surface layer at $x = 0$ and one at $x=L$, which will shortly prove convenient. The remainder of the problem now consists in determining $\alpha$, $\beta$, and the $a_k$ such that the boundary conditions (\ref{general-boundary-condition1}) and (\ref{general-boundary-condition2}) are satisfied. In general this is a difficult step in the analysis of two-way diffusion equations. One issue is that the eigenvalue problem in $\theta$, Eq. (\ref{eq:3-theta-eigenvalue-eqtn}), is not of the classical type since the weight function $\cos \theta$ is indefinite. Therefore, classical results from Sturm-Liouville theory regarding the completeness of eigenfunctions do not apply. Moreover, it is not obvious how to deal with the fact that the boundary conditions are split between $x=0$ and $x=L$.

For equations of our type, the first issue is not hard to address. It is possible to prove that the functions $\cos \theta$, $1$, and $\{\Theta_k\}_{k=-\infty}^{\infty}$ are complete on the range $(-\pi, \pi)$ \cite{Beals1981}. The expansion coefficients are easily determined from the orthogonality relations
\begin{align}
 \int_{-\pi}^{\pi} \Theta_k \cos \theta d\theta &= 0  \\
 \int_{-\pi}^{\pi} \Theta_k \cos^2 \theta d\theta &= 0, \hspace{8mm} k \neq 0 \\
 \int_{-\pi}^{\pi} \Theta_j \Theta_k \cos \theta d\theta &= \text{sgn}(j) \delta_{jk}, \hspace{8mm} (j,k) \neq (0,0),
\end{align}
obtained by directly integrating Eq. (\ref{eq:3-theta-eigenvalue-eqtn}) and assuming normalization $\int \Theta_k^2 \cos \theta d\theta = \text{sgn}(k)$. Next, this completeness result may be combined with an iterative procedure in order to address the issue of split boundary conditions. Referring to the boundary conditions in the form of Eqs. (\ref{general-boundary-condition1}) and (\ref{general-boundary-condition2}), we introduce the function
\begin{equation}
v(\theta)=\left\{
\begin{array}{cc}
v_{+}(\theta )-\alpha_0 +\beta_0 \cos \theta  & \cos \theta >0 \\
v_{-}(\theta )-\alpha_0 -\beta_0 \left( L-\cos \theta \right)  & \cos \theta <0
\end{array}
\right.
\end{equation}
and proceed as follows: First, we choose $\alpha_0$ and $\beta_0$ such that $\int v \cos \theta d\theta = 0 = \int v \cos^2 \theta d\theta$. Then, based on the orthogonality relations above, we can write $v(\theta) = \sum_{k \neq 0} a_k^0 \Theta_k$, where
\begin{equation}
a_k^0 = \text{sgn}(k) \int_{-\pi}^{\pi} v(\theta) \Theta_k \cos \theta d\theta.
\label{eq:3-coeff}
\end{equation}
Based on this construction, we have the identities
\begin{align}
v(\theta) &= \sum_{k>0}a_k^0\Theta_k+\sum_{k<0}a_k^0\Theta_k \\
               &= \sum_{k>0}a_k^0\Theta_k+\sum_{k<0}a_k^0e^{-\lambda _{k} L}\Theta_k + \sum_{k<0}a_k^0(1 - e^{-\lambda _{k} L})\Theta_k \\
               &= \sum_{k>0}a_k^0 e^{\lambda _{k} L}\Theta_k+\sum_{k<0}a_k^0\Theta_k + \sum_{k>0}a_k^0(1 - e^{\lambda _{k} L})\Theta_k.
\end{align}


In particular, restricting ourselves to the domain $\cos \theta > 0$, we can write
\begin{align}
v_{+}(\theta) &= \alpha_0 - \beta_0 \cos \theta + \sum_{k>0}a_k^0\Theta_k+\sum_{k<0}a_k^0e^{-\lambda _{k} L}\Theta_k + \sum_{k<0}a_k^0(1 - e^{-\lambda _{k} L})\Theta_k\\
&= f_0(0, \theta) + \sum_{k<0}a_k^0(1 - e^{-\lambda _{k} L})\Theta_k
\end{align}
where
\begin{equation}
f_0(x, \theta) = \alpha_0 + \beta_0 (x - \cos \theta) + \sum_{k>0}a_k^0e^{\lambda _{k}x}\Theta_k+\sum_{k<0}a_k^0e^{\lambda _{k}\left( x-L\right) }\Theta_k.
\end{equation}
Similarly, on the negative domain we have
\begin{align}
v_{-}(\theta) &= \alpha_0 + \beta_0(L - \cos \theta) + \sum_{k>0}a_k^0e^{\lambda _{k} L}\Theta_k+\sum_{k<0}a_k^0\Theta_k + \sum_{k>0}a_k^0(1 - e^{\lambda _{k} L})\Theta_k\\
&= f_0(L, \theta) + \sum_{k>0}a_k^0(1 - e^{\lambda _{k} L})\Theta_k.
\end{align}
Hence, $f_0$ is an approximate solution, with error given by
\begin{equation}
v_{err}^0=\left\{
\begin{array}{cc}
\sum_{k<0}a_k^0(1 - e^{-\lambda _{k} L})\Theta_k  & \cos \theta >0 \\
\sum_{k>0}a_k^0(1 - e^{\lambda _{k} L})\Theta_k  & \cos \theta <0.
\end{array}%
\right.
\end{equation}

We proceed in the natural way, applying the same procedure to $v_{err}^0$ as we did to $v(\theta)$, i.e. finding new coefficients $\alpha_1$, $\beta_1$, and $a_k^1$ such that
\begin{align}
&f_1 + v_{err}^1 = v_{err}^0
\end{align}
where
\begin{align}
&f_1 = \alpha_1 + \beta_1 (x - \cos \theta) + \sum_{k>0}a_k^1e^{\lambda _{k}x}\Theta_k+\sum_{k<0}a_k^1e^{\lambda _{k}\left( x-L\right) }\Theta_k \\
&v_{err}^1=\left\{
\begin{array}{cc}
\sum_{k<0}a_k^1(1 - e^{-\lambda _{k} L})\Theta_k  & \cos \theta >0 \\
\sum_{k>0}a_k^1(1 - e^{\lambda _{k} L})\Theta_k  & \cos \theta <0,
\end{array}%
\right.
\end{align}
and adding $f_1$ to the zeroth order solution $f_0$. The quantity $f_0 + f_1$ is then an improved estimate of the solution, with error given by $v_{err}^1$. Continuing in this way, at the $n^{th}$ step we obtain new corrections to the coefficients, which we denote by $\alpha_n$, $\beta_n$, and $a_k^n$. If the iterative procedure converges, the coefficients in the expansion (\ref{eq:3-two-way-sol-expansion}) are given by $\alpha = \sum_{n=0}^{\infty} \alpha_n$, $\beta = \sum_{n=0}^{\infty} \beta_n$, and $a_k = \sum_{n=0}^{\infty} a_k^n$.

\begin{figure}
  \includegraphics[width=0.35\linewidth,height=0.2038043478\linewidth]{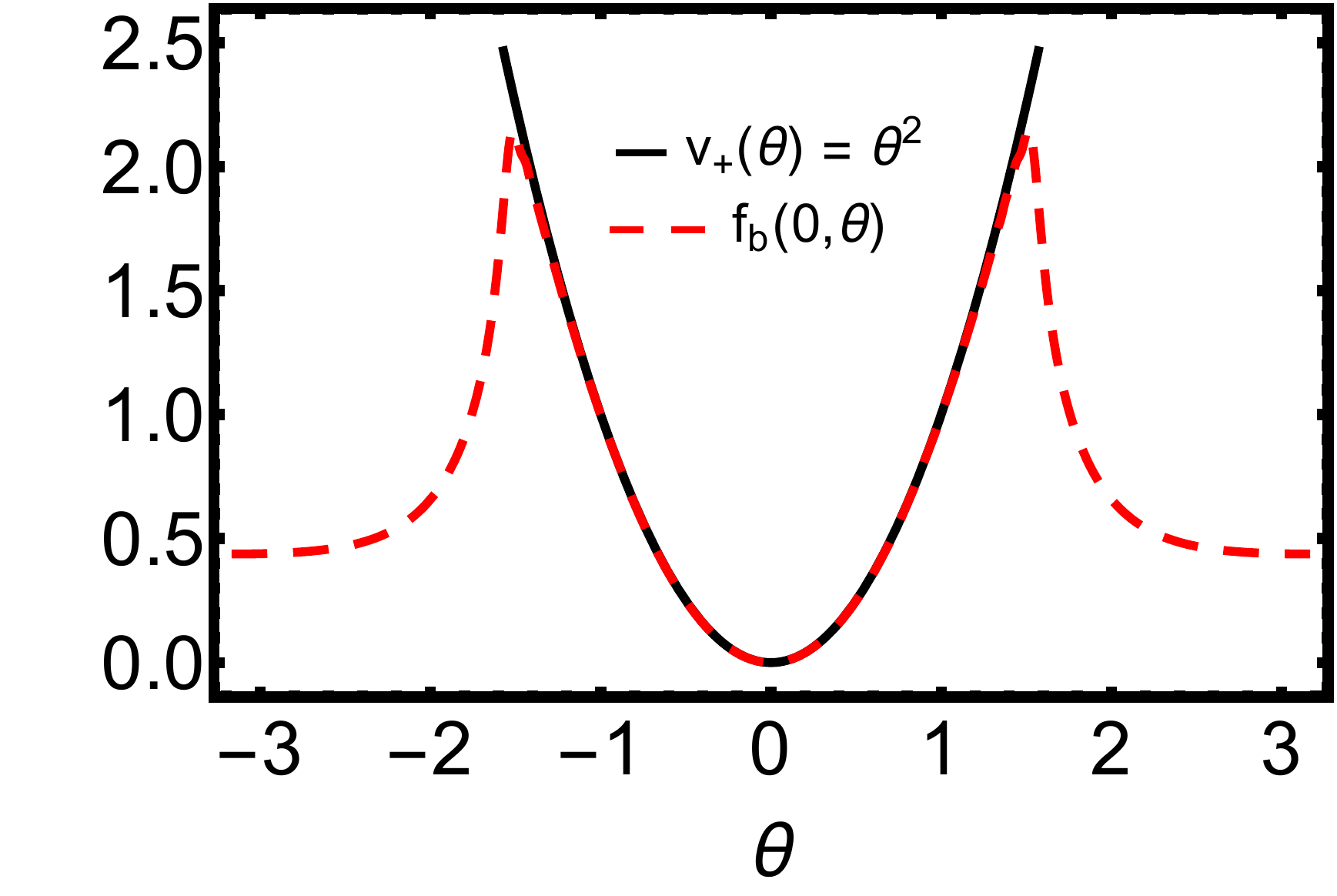}
\includegraphics[width=0.35\linewidth,height=0.2038043478\linewidth]{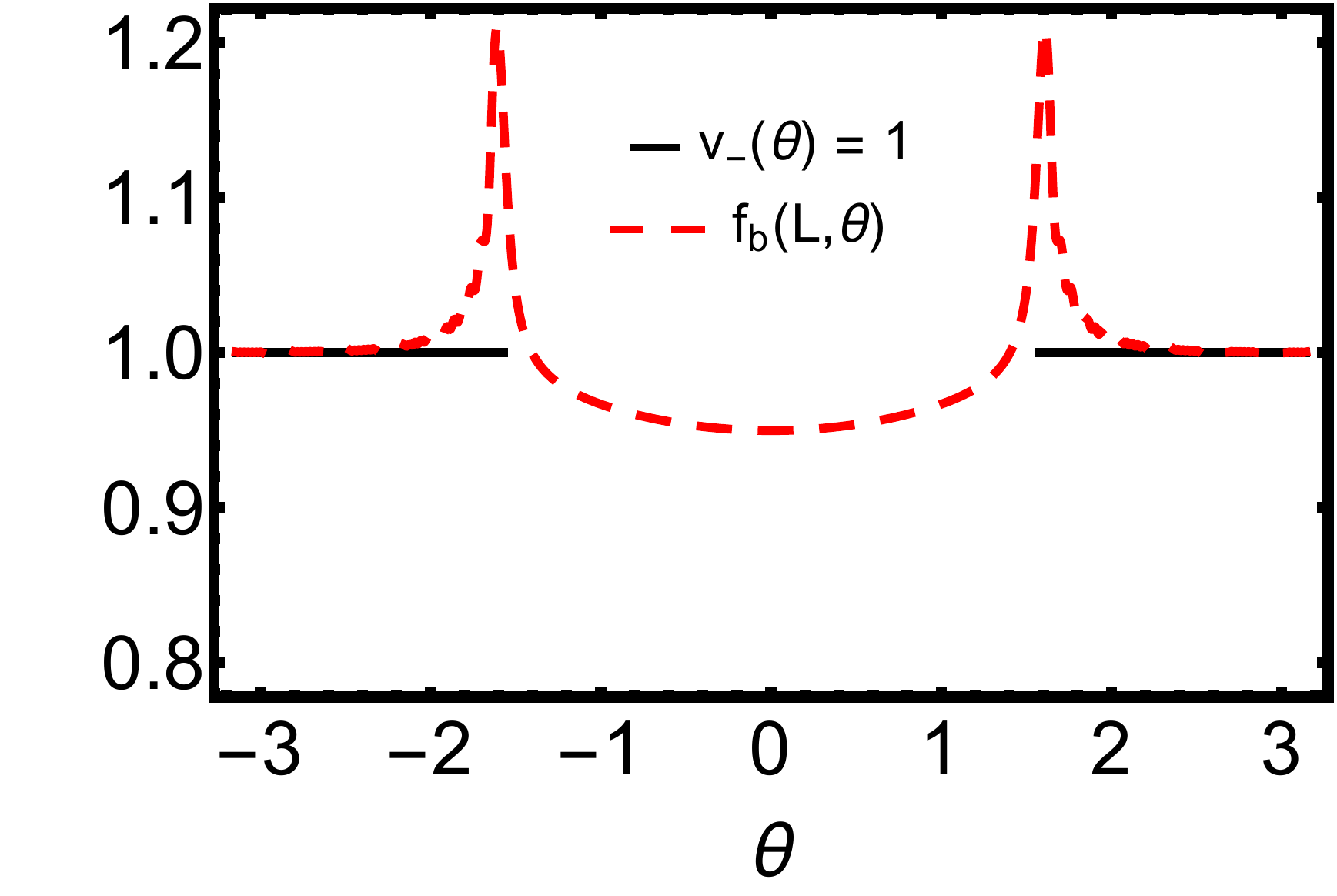}
  \includegraphics[width=0.35\linewidth,height=0.2038043478\linewidth]{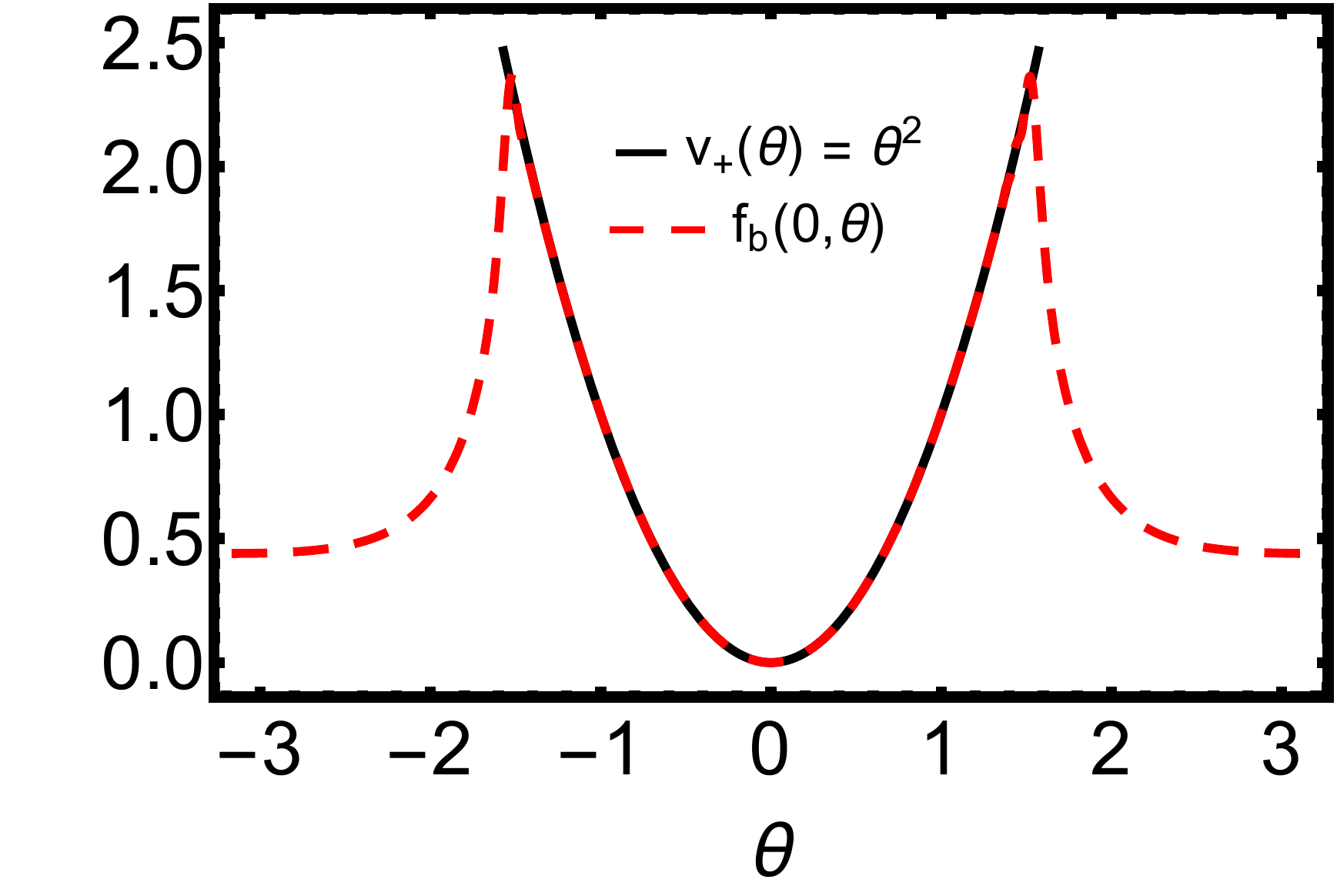}
\includegraphics[width=0.35\linewidth,height=0.2038043478\linewidth]{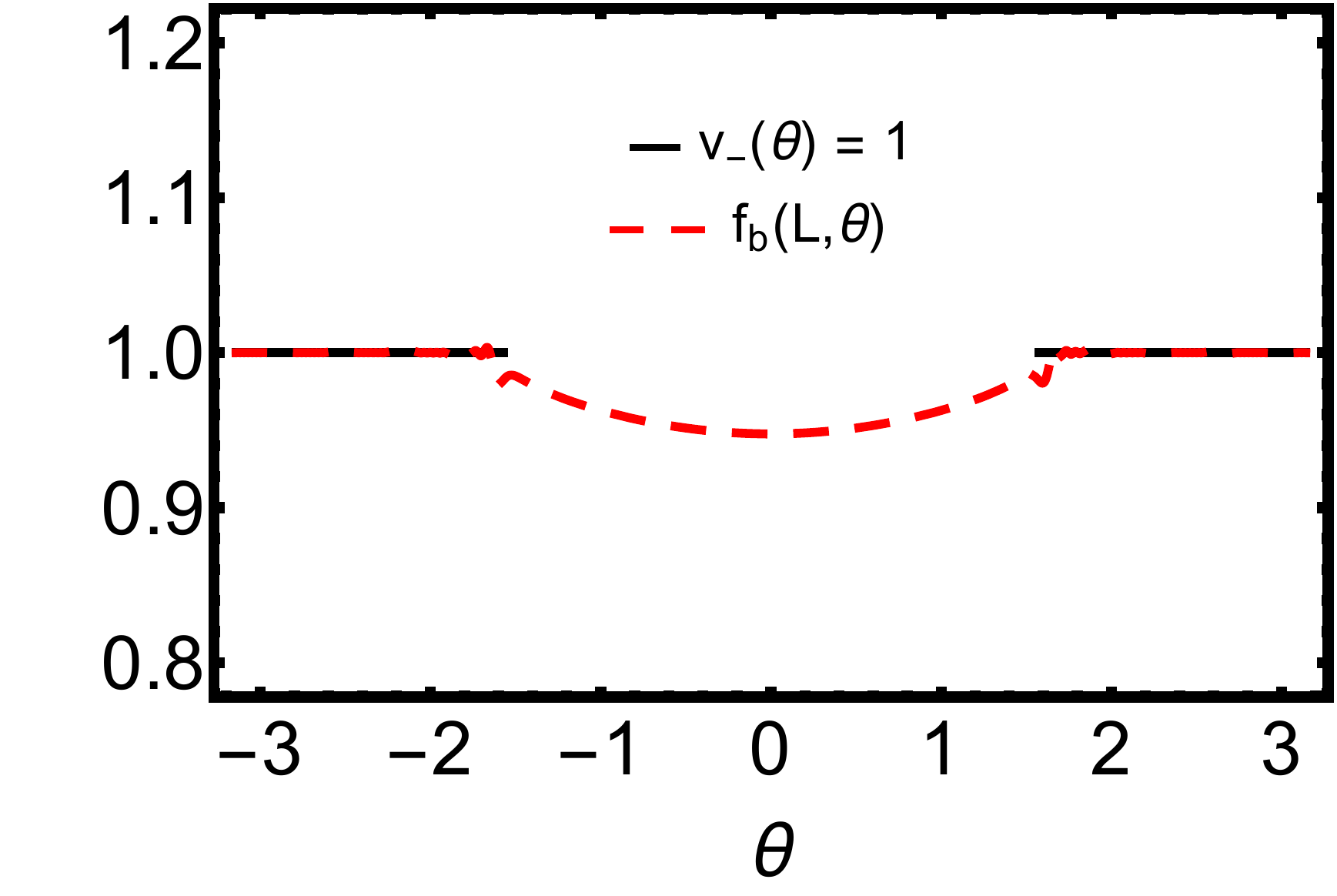}
  \caption{The iterative procedure seeks a function $f(x,\theta)$ expanded in the form of Eq. (\ref{eq:3-two-way-sol-expansion}) which satisfies the half-range boundary conditions $f(0,\theta) = v_{+}(\theta) $ where $\cos \theta > 0$, and $f(L,\theta) = v_{-}(\theta) $ where $\cos \theta < 0$. Here we show the results of this calculation for simple test functions $v_{+}(\theta) = \theta^2$ and $v_{-}(\theta) = 1$. The quantities $f(0,\theta)$ and $f(L,\theta)$ are dashed red, and $v_{\pm}(\theta)$ is solid black. The top two plots show the results for 5 iterations, and the bottom two plots for 100 iterations. In the latter case the error in $f(x,\theta)$ is nearly imperceptible.}
  \label{fig:iter_demonstration}
\end{figure}

In fact, the procedure is expected to converge since at each iteration, the $\Theta_k$ in $v_{err}^n$ are evaluated only on the range where they decay exponentially, in contrast to the dual range where they are large and oscillatory (see Fig. \ref{fig:eigenfunctions}). Hence, with respect to a suitably defined norm, $v_{err}^n$ is expected to go to $0$ for an arbitrarily large number of iterations. These assertions are supported by numerical calculation of the iterative procedure for various functions $v_{\pm}(\theta)$. For instance, Fig. \ref{fig:iter_demonstration} shows the results of applying the iterative procedure to determine $f(x,\theta)$ given boundary data of the form $v_{+}(\theta) = \theta^2$ and $v_{-}(\theta) = 1$ and taking $L$ to be nominally $20$. Calculating just 5 iterations (using 200 eigenfunctions) shows reasonable convergence, whereas after 100 iterations the discrepancy is nearly imperceptible. Thus, the iterative procedure enables order-by-order calculation of the full solution $f(x, \theta)$ to any desired accuracy.

\section{Confinement in a channel}

\begin{figure}
  \includegraphics[width=0.3640500000\linewidth,height=.45\linewidth]{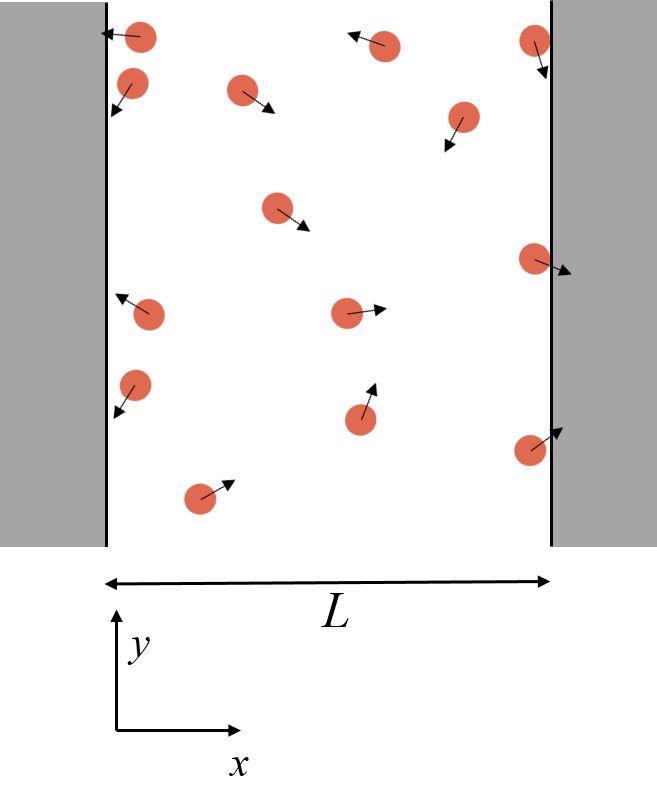}
  \caption{The infinite channel geometry for ABPs. The system is infinite in the $y$ direction and of uniform width $L$ in the $x$ direction. The boundaries at $x = 0$ and $x=L$ are impenetrable hard walls which enforce $0$ net flux in the $x$ direction, but otherwise do not impede the particles' rotation or motion in the $y$ direction.}
  \label{fig:infinite_channel}
\end{figure}

We are now in a position to formulate the problem of ABPs confined in a channel with hard walls (Fig. \ref{fig:infinite_channel}) and solve for the corresponding steady-state distribution. The channel is two-dimensional, of fixed width $L$ in the $x$ direction, and of infinite extension in the $\pm y$ direction. Although nominally two-dimensional, this system can be given an effective one-dimensional description due to the translational invariance in the $y$ direction. Moreover, the hard wall boundaries provide a strong demonstration of one of the unique characteristics of active matter under confinement, namely accumulation of particles at the boundary. This conclusion follows from a simple qualitative argument. In the bulk each particle spends an infinitesimal time $dx / v_0$ in the volume $dx$, whereas in the infinitesimal volume at the wall a particle spends a \emph{finite} time (on average) equal to $1 / D_r$. Therefore, the quantity $f(\boldsymbol{r}, \theta)$, which is an average over all particle trajectories, diverges at the boundary. The exact fraction of particles which accumulate at the wall (as a function of channel width $L$) is a quantity we will shortly calculate.

Our goal is to formulate the problem so that the boundary conditions take the form
\begin{eqnarray}
f(0,\theta ) &=&v_{+}(\theta ),\text{ \ \ \ \ \ where } \cos(\theta) >0 \\
f(L,\theta ) &=&v_{-}(\theta ),\text{ \ \ \ \ \ where } \cos(\theta) <0,
\end{eqnarray}
in which case the techniques from the previous section may be directly applied. In a confined geometry, however, these boundary conditions are not fixed from the outset, but instead determined self-consistently from the distribution of particles adsorbed onto at the wall. It is therefore useful to treat the dynamics in the bulk separately from the dynamics at the wall and couple the two. We note that Lee \cite{Lee2013} has formulated the problem in similar terms. To this end, we consider a steady-state distribution in the bulk, $f_b(\boldsymbol{r}, \theta)$, and a steady-state distribution at the wall, $f_w(\theta)$, and proceed to write down the defining equations for both.


\textbf{Description in the bulk}

Due to translational invariance in the vertical direction, $f_b$ depends only on $x$ and $\theta$, and satisfies
\begin{equation}
\ell_p \cos \theta \frac{\partial f_b}{\partial x}=\frac{\partial ^{2}f_b}{\partial \theta ^{2}},
\label{eq:4-bulk1}
\end{equation}
which is the same as Eq. (\ref{FP-1d-dimensional}). The boundary conditions in this case are given by two delta function sources at each wall. Intuitively, one can imagine a particle leaving the wall and entering the bulk as soon as its orientation points away from the wall. The rate at which this reorientation occurs is
\begin{equation}
J = D_{r}\left\vert \frac{d f_w}{d \theta }\right\vert _{\theta =\pm \pi /2} \delta(\theta \pm \pi / 2)
\end{equation}
in units of particles per unit time per unit angle. Dividing this flux $J$ by the particle velocity $\pm v_0 \cos \theta$, we identify the boundary conditions for Eq. (\ref{eq:4-bulk1}) at the walls as

\begin{align}
f_b(0, \cos \theta > 0) = \lim_{\theta' \rightarrow \pi /2^{-}} \frac{1}{\ell_p \cos\theta} \left\vert \frac{d f_w}{d \theta }\right\vert _{\theta = \theta'} \delta(\theta - \theta') +\lim_{\theta' \rightarrow -\pi /2^{+}} \frac{1}{\ell_p\cos\theta} \left\vert \frac{d f_w}{d \theta }\right\vert _{\theta = \theta'} \delta(\theta + \theta')
\label{eq:4-bulk2}
\end{align}
and
\begin{align}
f_b(L,\cos \theta < 0) = \lim_{\theta' \rightarrow \pi /2^{+}} \frac{1}{-\ell_p \cos\theta} \left\vert \frac{d f_w}{d \theta }\right\vert _{\theta = \theta'} \delta(\theta - \theta') +\lim_{\theta' \rightarrow -\pi /2^{-}} \frac{1}{-\ell_p\cos\theta} \left\vert \frac{d f_w}{d \theta }\right\vert _{\theta = \theta'} \delta(\theta + \theta').
\label{eq:4-bulk3}
\end{align}
 which matches the general form sought, Eqs. (\ref{general-boundary-condition1}) and (\ref{general-boundary-condition2}). While the boundary conditions as written appear irreparably singular, this issue does not arise in practice. In a physical system the orientational fluctuation that causes a particle on the wall to turn into the bulk will always be finite, such that the above terms are evaluated very near, but not exactly at $\pm \pi / 2$. This is also the approach adopted for all analytical calculations: We choose $\theta'$ to be as close to $\pm \pi / 2$ as necessary to give accurate results.

\textbf{Description at the wall}

The dynamics at the wall is dictated by the flux of particles incoming from the bulk. Picturing for the moment these particles coming in discrete waves of duration $\Delta t$, each wave (at $x = L$, for instance) corresponds to an initial angular distribution on the wall given by
\begin{equation}
g(\theta, t = 0) = v_0 \Delta t \cos \theta f_b(L, \theta), \hspace{8mm} \cos \theta > 0,
\end{equation}
 which evolves via the angular diffusion equation
\begin{equation}
\frac{\partial g(\theta,t)}{\partial t}=D_{r}\frac{\partial ^{2}g(\theta,t)}{\partial \theta ^{2}}
\label{boundary_diffusion}
\end{equation}
with absorbing boundaries at $\theta = \pm \pi / 2$. The steady-state density at the wall is a sum over these profiles, in the limit $t \rightarrow \infty$ and $\Delta t \rightarrow 0$. If we assume the first wave comes at $t = 0$ (the result is independent of the choice of origin), we find
\begin{equation}
f_{w}(\theta )=\int_{0}^{\infty } g\left( \theta,t\right) dt
\label{wall1}
\end{equation}
where $g\left( \theta,t\right)$ satisfies Eq. (\ref{boundary_diffusion}) with initial condition
\begin{equation}
g(\theta, t = 0) = v_0 \cos \theta f_b(L, \theta), \hspace{8mm} \cos \theta > 0.
\label{wall2}
\end{equation}
Analogous statements hold for the wall at $x = 0$. In an equivalent and slightly more direct formulation, we could also determine $f_w$ by implicitly including the impinging flux as a source term in the steady-state equation for $f_w$. For example, at $x=L$ we would have
\begin{equation}
\frac{d^2 f_w}{d \theta^2} = \ell_p \cos \theta f_b(L, \theta), \hspace{8mm} \cos \theta > 0.
\end{equation}

In summary, we have now reduced our problem to solving for $f_b$ and $f_w$ in the preceding coupled equations. At first glance this appears difficult, since in general $f_b$ is needed to determine $f_w$, and vice versa. However, the symmetry of the problem allows a considerable simplification. We denote $J^L_{in}$ and $J^L_{out}$ as the rates at which particles enter and leave the bulk at $x = 0$, and $J^R_{in}$ and $J^R_{out}$ as the rates at $x = L$. Due to the symmetry of the boundaries, $J^L_{in} = J^R_{in}$ and $J^L_{out} = J^R_{out}$. Moreover, since the distribution in the bulk is solved in steady state, $J^L_{in} + J^R_{in} = J^L_{out} + J^R_{out}$. Combining the two, we see that all fluxes are equal, which implies that the steady-state condition at the walls is automatically enforced. If this is true, we may solve for $f_b$ assuming some arbitrary value of $\left\vert \frac{d f_w}{d \theta }\right\vert _{\theta = \pm \pi / 2}$, subsequently use this particular $f_b$ to find $f_w$, and be guaranteed that this $f_w$ is consistent with the initial assumption. The quantity $\left\vert \frac{d f_w}{d \theta}\right\vert _{\theta = \pm \pi / 2}$ then simply functions as a normalization constant in front of the entire solution.

\textbf{Signatures of confinement}

\emph{Calculation of the expansion coefficients.} Having formulated the boundary conditions in the proper form, we are now prepared to calculate the expansion for the general solution of Eq. (\ref{eq:4-bulk1}):
\begin{equation}
f_b(x, \theta) = \alpha + \beta (x - \cos \theta) + \sum_{k>0}a_{k}e^{\lambda _{k}x}\Theta_k+\sum_{k<0}a_{k}e^{\lambda _{k}\left( x-L\right) }\Theta_k,
\end{equation}
where again $\ell_p$ has been set to 1 for notational convenience.
The first step in the iterative procedure for the expansion coefficients is to solve for $\alpha_0$ and $\beta_0$. Defining
\begin{equation}
\widetilde{v} = \left\{
\begin{array}{cc}
v_{+}(\theta )  & \cos \theta >0 \\
v_{-}(\theta )  & \cos \theta <0%
\end{array}%
\right.,
\end{equation}
 $I_1 = \int \widetilde{v} \cos \theta d\theta$, and $I_2 = \int \widetilde{v} \cos^2 \theta d\theta$ , we find that
\begin{align}
\alpha_0 &= \frac{L}{2(2L+\pi)} I_1 + \frac{I_2}{\pi} \\
\beta_0 &= -\frac{1}{2L + \pi} I_1.
\end{align}

From symmetry considerations, $I_1 = 0$. Moreover, this remains true at any step of the iterative procedure, such that $\beta$ is identically 0 and, referring to Eq. (\ref{eq:3-coeff}), $a_k = a_{-k}$. This also implies that $\alpha$ and the $a_k$ are independent of $L$ to order $e^{-\lambda^*L}$, where $\lambda^* \simeq 10.7$ is the smallest positive eigenvalue of the contributing $\Theta_k$. Thus, if we carry out the iterative procedure to arbitrary order, the solution takes the form
\begin{equation}
f_b(x, \theta) = A \left[ \alpha + \sum_{k>0}a_{k}e^{\lambda _{k}x}\Theta_k+\sum_{k<0}a_{-k}e^{\lambda _{k}\left( x-L\right) }\Theta_k \right].
\label{eq:4-fb-expansion}
\end{equation}
Here $A$ is a normalization constant which depends on $L$, but $\alpha$ and the $a_k$ are independent of $L$ to order $e^{-\lambda^*L}$; their numerical values can be calculated using the iterative procedure. Using this as a starting point, it is possible to derive a variety of interesting results.

\begin{figure}
  \includegraphics[width=0.38\linewidth,height=.28\linewidth]{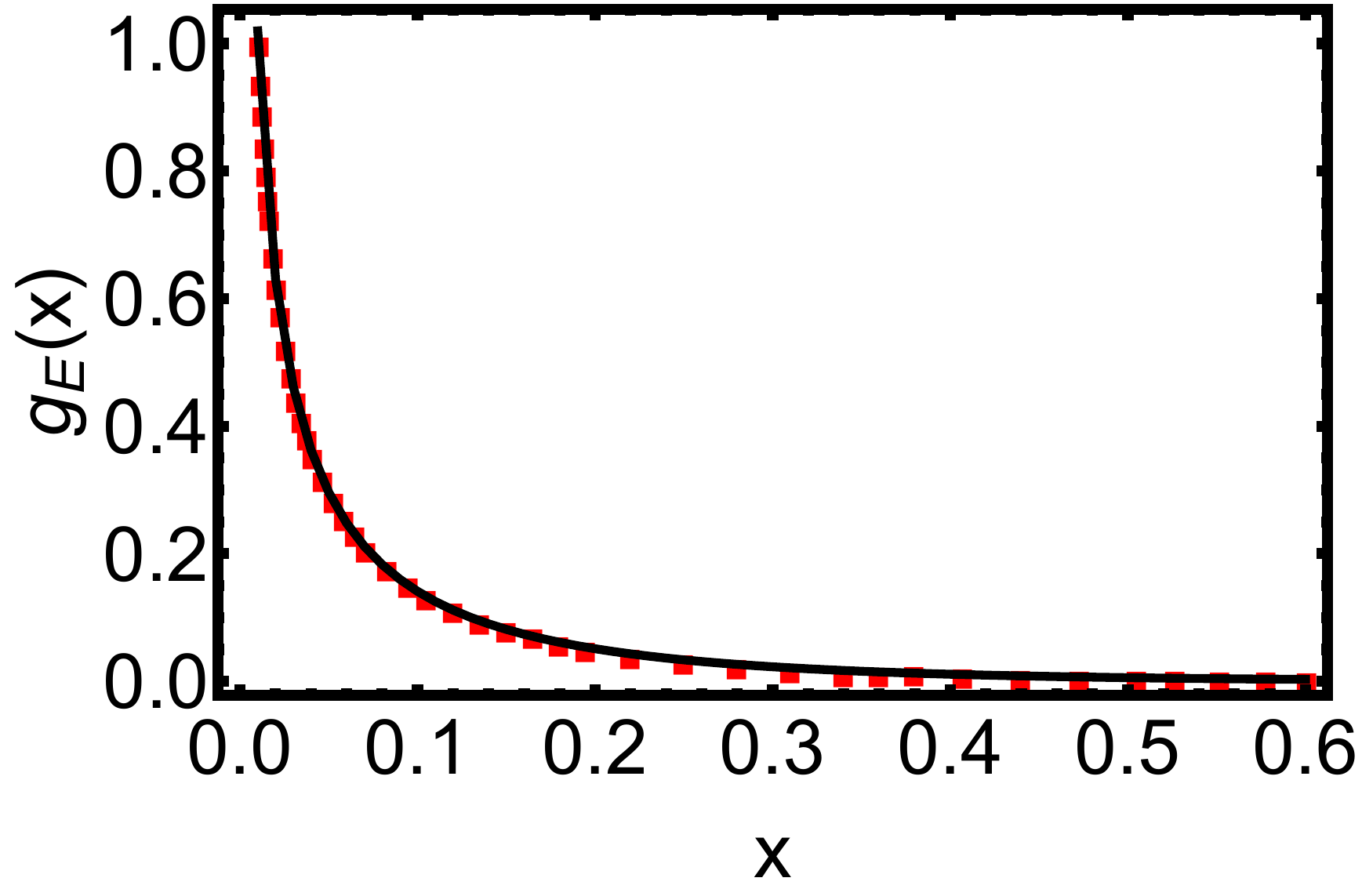}
  \caption{Scaling of the excess density $g_E(x)$ near $x = 0$ and for $L \gg 1/\lambda^*$ as measured in simulations (red squares), and using the fit in Eq. (\ref{reduced-density-power-law}), with $a \simeq 0.134$. The normalization chosen is arbitrary. The simulated channel width is $L=7$, $\lambda^*\simeq 10.7$, and all other simulation details are given in Appendix C.}
  \label{fig:bulk_dist}
\end{figure}

\emph{Density in the bulk.} From Eq. (\ref{eq:4-fb-expansion}) we see that the density $g(x) = \int f_b(x, \theta) d\theta$ is large near the walls but decays rapidly to the constant $2\pi A \alpha$ in the bulk. Far from the wall, the solution is dominated by a single exponential decay. Near the wall, however, the cumulative effect of the separable solutions cannot be ignored. In fact, as explained in Appendix B, the solution displays a power-law divergence as the wall is approached. More precisely, we may approximate the exact expansion (when $L \gg 1/\lambda^*$) using
\begin{equation}
g_E(x) \sim \frac{a}{\sqrt{x}}\left[1 - \text{erf}(2.17\sqrt{x})\right]
\label{reduced-density-power-law}
\end{equation}
where $x$ is the distance from the left wall, $a$ is a constant, and $g_{E}(x)$ is the excess density defined as $g_{E}(x) = \int f_b(x,\theta) d\theta - 2 \pi A \alpha$. This trend matches simulation data, as shown in Fig. \ref{fig:bulk_dist}.

This analysis additionally tells us that the influence of the walls disappears for locations in the channel greater than $\sim \ell_p / 2$ units from either wall. This is expected to be a generic result for active systems; in fact, a similar boundary layer has been observed in run-and-tumble particles, with the boundary layer thickness scaling as the run length in place of $\ell_p$ \cite{Ezhilan2015}. We emphasize that for active colloids, as opposed to thermal colloids, such density variations are often non-negligible since $\ell_p$ can be comparable to system size \cite{Nash2010,Fily2014,Elgeti2013}.

\emph{Orientational order in the bulk.}
Following the previous discussion, we now show that a local increase in spatial density is not the only effect of boundaries. In line with the work of Enculescu and Stark \cite{Enculescu2011}, which demonstrated that torque-free external potentials can induce orientational ordering in active systems, we find that the presence of a boundary is sufficient to induce orientational order which extends a distance of $\sim \ell_p / 2$ into the bulk. To make this statement precise, we define a polarization field $\boldsymbol{P}(x)$ and nematic alignment field $\boldsymbol{Q}(x)$ by
\begin{align}
\boldsymbol{P} &= \int \hat{\boldsymbol{\nu}} f_b(x,\theta) d\theta \\
Q_{ij} &= \int \left(\hat{\nu}_i\hat{\nu}_j - \frac{1}{2} \delta_{ij}\right) f_b(x,\theta) d\theta
\end{align}
where again $\hat{\boldsymbol{\nu}} = (\cos \theta, \sin \theta)$. Because of the confinement in the $x$-direction, there cannot be a net flow of particles to the left or right. Therefore, the $x$ component of the polarization is $0$, as can be verified from Eq. (\ref{eq:4-fb-expansion}). Moreover, since $f_b(x,\theta)$ is an even function of $\theta$, the $y$ component is also $0$. Thus, there is no polar order. On the other hand, $\boldsymbol{Q}(x)$ is nonzero, and given by
\begin{equation}
\boldsymbol{Q} = \frac{1}{2}\left(
\begin{array}{cc}
-g_{E}(x) & 0 \\
0 & g_{E}(x)%
\end{array}%
\right).
\end{equation}
Since $g_E(x)$ is everywhere positive, particles prefer to align in the $\pm y$ direction. Moreover, this tendency is strongest near the walls, and for large enough $L$ decays via Eq. (\ref{reduced-density-power-law}) to (nearly) $0$ in the middle of the channel, where the orientations are approximately isotropic.



\begin{figure}
  \includegraphics[width=0.4347826087\linewidth,height=.3260869565\linewidth]{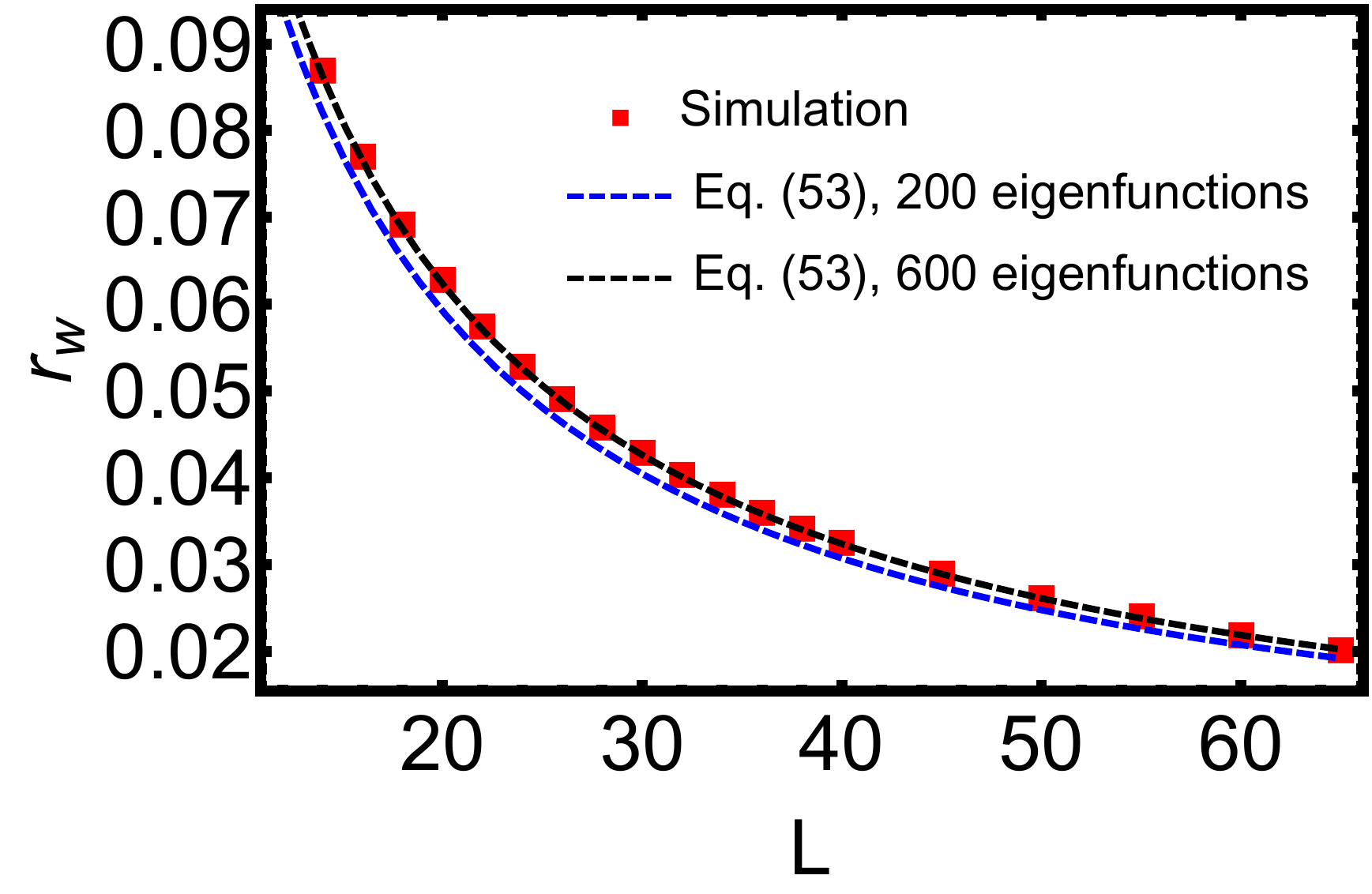}
\includegraphics[width=0.4347826087\linewidth,height=.3260869565\linewidth]{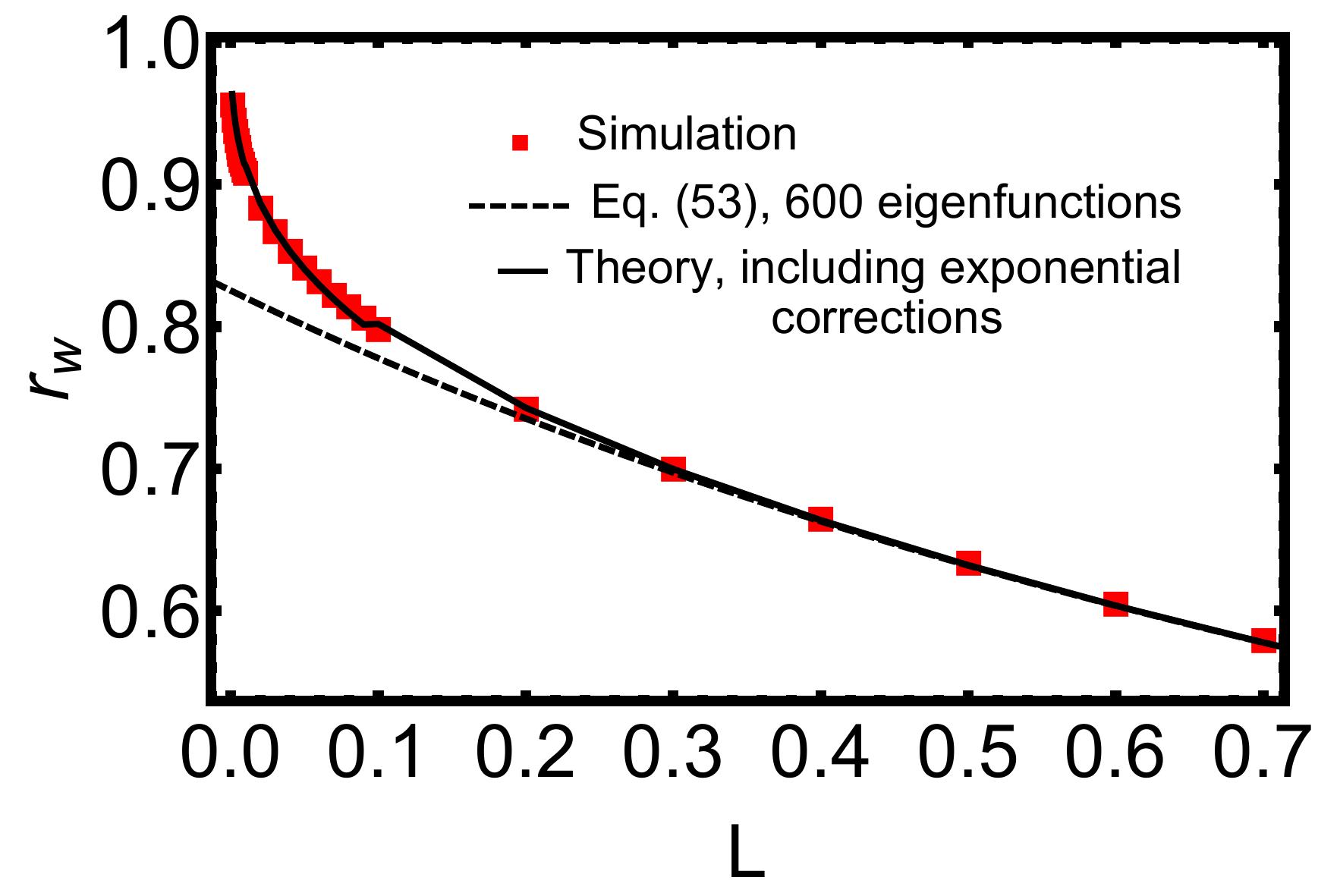}
  \caption{Fraction of particles accumulated at the wall. Left: Wall fraction scaling from simulations (red squares), and as predicted from Eq. (\ref{wall-fraction-scaling}) using 200 eigenfunctions (dashed blue) and 600 eigenfunctions (dashed black). For the latter calculation, the iterative procedure predicts $c_1 \simeq 1.349$ and $c_2 \simeq 1.635$. We note that because the simulation timestep $\Delta t$ is finite, there is a small spread of order $\epsilon \sim \sqrt{2 D_r \Delta t}$ in the initial angles of particles entering the bulk. To approximate this effect, the theory calculation takes $\theta' = \pm \pi / 2 \mp \epsilon$ in Eq. (\ref{eq:4-bulk2}) and $\theta' = \pm \pi / 2 \pm \epsilon$ in Eq. (\ref{eq:4-bulk3}). Right: Wall fraction for small $L$. The scaling from Eq. (\ref{wall-fraction-scaling}) using 600 eigenfunctions (dashed black) is compared with the simulation results (red squares) and the calculation from the iterative procedure keeping all the exponential corrections (solid black). As predicted, the scaling given by Eq. (\ref{wall-fraction-scaling}) breaks down near $L \approx 1/\lambda^{*} \approx 0.0939$.}
  \label{fig:wall_fraction}
\end{figure}

\emph{Scaling relations.} The form of the solution (\ref{eq:4-fb-expansion}) also allows the derivation of general scaling laws for the bulk density and the fraction of particles at the wall. Integrating both sides over the bulk, we find
\begin{equation}
r_b = A \left[ 2 \pi L \alpha + 2 C \right] + \mathcal{O}(e^{-\lambda^*L})
\end{equation}
where $r_b$ is the fraction of particles in the bulk, $C =  \int_0^{\infty} \int_{-\pi}^{\pi} \left[ \sum_{k>0}a_{k}e^{\lambda _{k}x} \Theta_k \right] d\theta dx$, and $\lambda^*$ is again the smallest positive eigenvalue of the contributing $\Theta_k$.
Next, we notice that because of the equations which couple $f_w$ to $f_b$, and the fact that the only piece of $f_b(0,\theta)$ and $f_b(L,\theta)$ which depends on $L$ to this order is $A$, $f_w$ itself will be proportional to $A$ and otherwise independent of $L$. That is, if $r_w$ is the fraction of particles at the wall, then for some constant $E$ independent of $L$,
\begin{equation}
r_w = E A + \mathcal{O}(e^{-\lambda^*L}).
\end{equation}
Finally, by normalization $r_b + r_w = 1$. Solving these equations for $r_w$ gives
\begin{equation}
r_w = \frac{c_1}{L + c_2} + \mathcal{O}(e^{-\lambda^* L})
\label{wall-fraction-scaling}
\end{equation}
for constants $c_1, c_2$ given in terms of $E$, $C$, and $\alpha$. These constants can either be fit from data or calculated directly from the iterative procedure, showing good agreement with simulation (Fig. \ref{fig:wall_fraction}).
We can use a similar procedure to show that the bulk density $g_b \equiv 2 \pi A \alpha$ scales as
\begin{equation}
g_b = \frac{1}{L + c_3} + \mathcal{O}(e^{-\lambda^* L}).
\end{equation}
We note that a similar scaling relation for $r_w$ has been obtained previously for active systems, for instance using a large $L$ approximation or dimensional analysis \cite{Elgeti2013,Elgeti2015}. Because the approach used here is exact, however, we are also able to identify the region in which the scaling relation breaks down. Based on the above analysis, this should occur near $L \approx 1/\lambda^{*} \approx 0.0939$, which agrees well with simulation results shown in Fig. \ref{fig:wall_fraction}.

An important physical outcome of these scaling relations is a clear violation of extensivity for narrow channels. For instance, since $f_w(\theta)$ is independent of $L$ (to order $\mathcal{O}(e^{-\lambda^* L})$), we have the following relation for the wall pressure $P_w$:
\begin{equation}
P_w \propto \frac{c_1}{L+c_2}
\end{equation}
which applies for varying $L$ and fixed particle number.
For large enough $L$, this reduces to $P_w \propto 1/L$, as expected for an extensive system. For narrow channels, however, $P_w$ is no longer proportional to $1/L$, and the actual pressure is in fact smaller than predicted by an extensivity argument. This reduction of pressure in narrow channels has been reported elsewhere on the basis of simulations and approximate analytic results \cite{Speck2016,Yan2015,Ezhilan2015}.

\begin{figure}
  \includegraphics[width=0.4\linewidth,height=.29\linewidth]{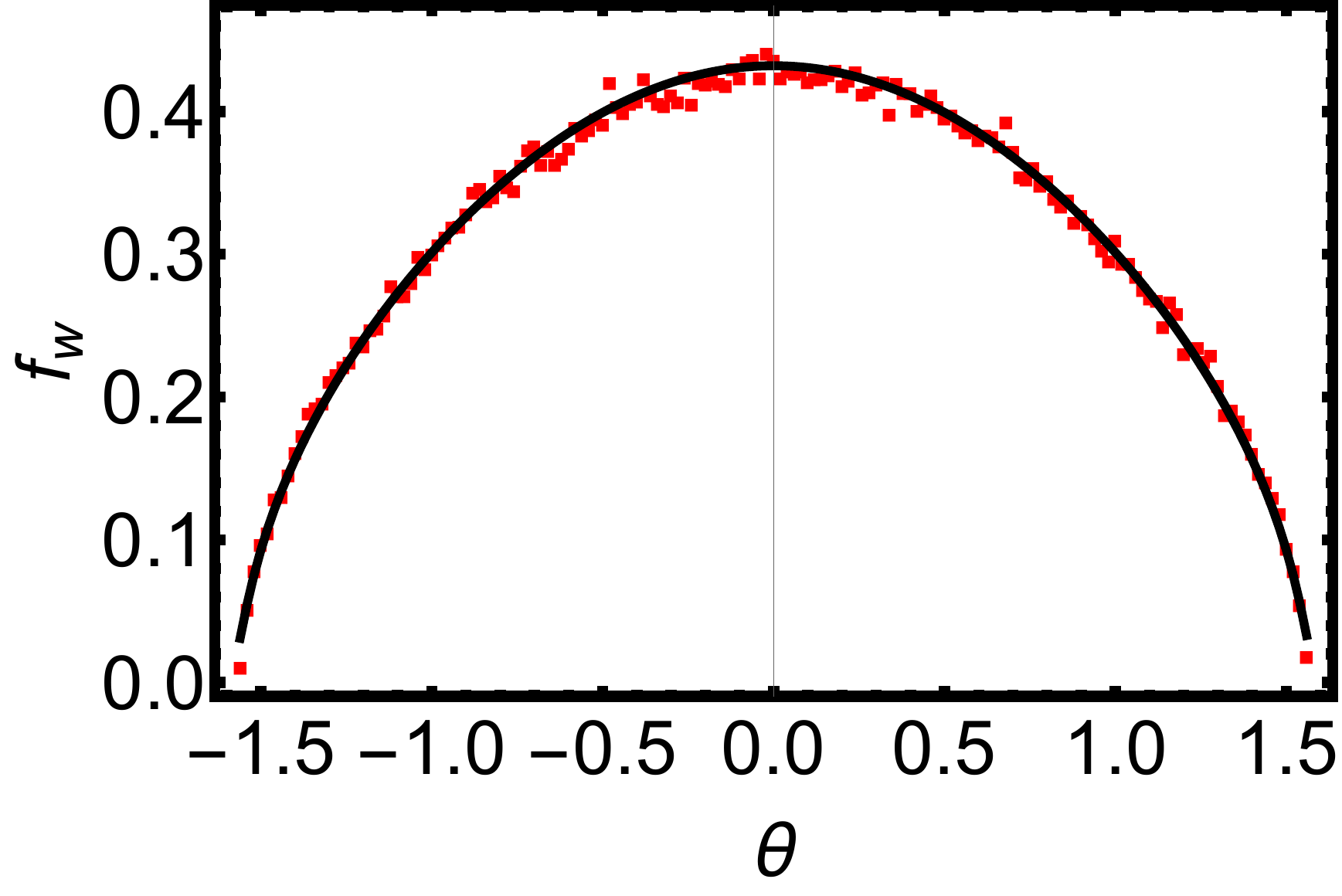}
  \caption{Wall distribution for $L = 20$ from simulations (red squares) and theory (black).}
  \label{fig:wall_dist}
\end{figure}

If we wish to quantitatively calculate the pressure, then we must know the distribution at the wall, which can be computed here using the full iterative procedure. Simulations confirm that the shape of the wall distribution $f_w(\theta)$ is independent of $L$ to order $\mathcal{O}(e^{-\lambda^* L})$, as predicted by the theory. The quantitative agreement between the full iterative procedure and the simulated $f_w(\theta)$ is also good, as shown in Fig. \ref{fig:wall_dist}. In particular, calculating the moment $\int_{-\pi/2}^{\pi/2} \cos \theta f_w(\theta) d\theta \simeq 0.739$ (where $f_w$ is normalized to 1) relates the number of particles at the wall to the pressure they exert.


\section{Constant flux steady state}

To study a constant flux steady state, we replace the channel walls at $x = 0$ and $x = L$ with reservoirs of ABPs having uniform (independent of $\boldsymbol{r}$ and $\theta$) distributions $\rho_L$ and $\rho_R$, respectively. In this case $v_+(\theta) = \rho_L$ and $v_-(\theta) = \rho_R$. (Note that to maintain reservoirs with this property in a real-life system, it may be necessary to remove particles as they leave the channel.) The interesting piece of this calculation is the diffusion solution $\beta (x - \cos \theta)$. As discussed in Section III, this solution cannot be expressed as a sum of separable solutions, and its inclusion is necessary for general boundary conditions. Physically, it accounts for nonzero flux: The constant $\beta$ is directly proportional to the net flux in the channel. In particular, dividing this flux by the gradient $(\rho_R - \rho_L)/L$ can supply information about the effective diffusivity of ideal ABPs. Let us see how this works out in detail; in what follows we derive $\beta$ to first order in the iterative procedure.

To zeroth order,
\begin{equation}
\beta_0 = \frac{2}{2L + \pi}\left(\rho_R - \rho_L\right).
\end{equation}
The coefficients $a_k^0$ may be calculated in the usual way. They are
\begin{equation}
a_k^0 = \left(\rho_L - \rho_R + \beta_0 L \right) X_k
\end{equation}
where $X_k = \text{sgn}(k) \int_{\cos \theta > 0} \Theta_k \cos \theta d\theta$. From $\beta_0$ and $a_k^0$ we may then calculate $\beta_1$, with the result
\begin{align}
\beta_1 =& -2Z\frac{\rho_L - \rho_R + \beta_0 L}{2L + \pi} \\
Z \equiv& \int_{\cos \theta > 0} \left( \sum_{k<0} X_k \Theta_k \right) \cos \theta d\theta \approx -0.0699.
\end{align}
We note that the constant $Z$ is independent of system parameters. Putting these results together,
\begin{equation}
\beta \simeq \beta_0 + \beta_1 = \left[ \left(1 + Z \right) \frac{2}{2L + \pi} - Z L \left( \frac{2}{2L + \pi} \right)^2 \right] \left(\rho_R - \rho_L\right).
\label{eq:5-beta}
\end{equation}
Let us now define an effective diffusivity for the ideal ABPs:
\begin{equation}
D_A \equiv -\frac{J}{\nabla \rho} \\
\end{equation}
where $J = -\pi v_0 \beta$ is the flux inside the channel, and $\nabla \rho = \left(\rho_R - \rho_L\right) / L$. Substituting from Eq. (\ref{eq:5-beta}) gives
\begin{equation}
D_A \simeq \left(1 + Z \right) \frac{2 \pi v_0 L}{2L + \pi} - \pi v_0 Z \left( \frac{2 L}{2L + \pi} \right)^2.
\label{active-diffusivity}
\end{equation}
This result can be interpreted in light of the well-established relationship between mean-squared displacement and transport coefficients in 1d: Given $\langle \Delta x^2 \rangle \sim t^\alpha$ with $0 < \alpha \leq 2$, the diffusivity $D$ scales as $L^{2 - 2/\alpha}$ \cite{Baowen2003}. For instance, ordinary diffusion gives rise to a constant $D$ (Fick's law of diffusion), whereas ballistic trajectories result in $D \propto L$. For ABPs, however, the mean-squared displacement scales differently with $t$ depending on what time regime we are interested in: ballistic for times much shorter than the reorientation time and diffusive for times much longer. Therefore, the result in Eq. (\ref{active-diffusivity}) may be seen as a generalization of the scaling law $L^{2 - 2/\alpha}$ for a mean-squared displacement that cannot be written in terms of a simple power of $t$. Examining Eq. (\ref{active-diffusivity}), we see that $D_{A}$ does reduce to the appropriate limits: As $L \rightarrow 0$, ballistic motion dominates so that $D_A \propto L$, while as $L \rightarrow \infty$ diffusion dominates and $D_A \rightarrow \text{constant}$. Reinserting the factors of $\ell_p$, we find that this constant is proportional to $\frac{v_0^2}{D_r}$, a familiar result for ABPs \cite{Marchetti2016}.

\section{Sedimentation}

The analysis in the presence of a uniform external field can be approached using similar techniques. We assume equations of motion given by
\begin{align}
  \dot{\boldsymbol{r}} &= v_0 \hat{\boldsymbol{\nu}} + \textbf{\textrm{f}}
  \label{eq:6-abp-spatial-force}
  \\
  \dot{\theta} &= \sqrt{2 D_r} \eta^\text{R}
  \label{eq:6-abp-rotational-force}
\end{align}
where $\textbf{\textrm{f}} = -\nabla V(x) / \zeta = (-F / \zeta, 0)$ is an external force, independent of  $x$ and $t$, that drives particles toward a boundary at $x=0$, and $\zeta$ is the friction. The associated Smoluchowski equation is
\begin{equation}
(\cos \theta - r) \frac{\partial f}{
\partial x}=(1/\ell_p)\frac{
\partial ^{2}f}{\partial \theta ^{2}}
\label{eq:FP-1d-force}
\end{equation}
where $r = F/v_0 \zeta$ and $\ell_p = v_0/D_r$. Since the self-propulsion of ABPs has fixed norm, we also require $|r| < 1$ in order to avoid collapse of the distribution function under the external force. If we consider Eq. (\ref{eq:FP-1d-force}) on the finite interval $0 < x < L$, boundary data are specified as
\begin{eqnarray}
f(0,\theta ) &=&v_+(\theta),\text{ \ \ \ \ \ where } \cos(\theta) - r >0 \\
f(L,\theta ) &=&v_-(\theta),\text{ \ \ \ \ \ where } \cos(\theta) - r <0.
\end{eqnarray}

\textbf{Separation of variables}

Proceeding as before by separation of variables, the separable solutions $\Gamma(x) \Theta(\theta)$ now satisfy
\begin{eqnarray}
\frac{d\Gamma }{dx} &=&\frac{\lambda}{\ell_p}\Gamma  \\
\frac{d^{2}\Theta }{d\theta ^{2}} &=&\lambda (\cos \theta - r) \Theta, \hspace{8 mm} \text{with} \hspace{2 mm} \Theta(\theta) = \Theta(\theta + 2\pi).
\label{eq:6-theta-eigenvalue-eqtn-force}
\end{eqnarray}
The angular eigenfunctions can be constructed as in the zero force case using a Fourier expansion (see Appendix A). The major difference in the spectral structure is that a diffusion solution no longer exists, since this would require the existence of a function $G(\theta)$ satisfying
\begin{equation}
\frac{d^2 G}{d\theta^2} = \cos \theta - r
\end{equation}
as well as the periodic boundary conditions. However, integrating over $\theta$ and using continuity of $\frac{dG}{d\theta}$, we obtain $0 = -2 \pi r$. Hence, no diffusion exists for $r \neq 0$. On the other hand, there does exist an eigenvalue $\lambda_R = -2 r + \mathcal{O}(r^3)$ which merges with the $0$ eigenvalue as $r \rightarrow 0$. In this limit, the associated separable solution $e^{\lambda_{R}x/\ell_p} R(\theta)$ is linearly independent from the constant solution; by choosing the appropriate linear combination and normalization, the diffusion solution is recovered (see Section III in Ref. \cite{Kruskal1980} for a detailed demonstration in the context of a similar problem).  

Indexing the remaining eigenvalues and eigenfunctions with $k$ such that positive $k$ corresponds to the negative part of the spectrum, and vice versa, we can write the general expansion as
\begin{equation}
f(x, \theta) = \alpha + \beta e^{\lambda_{R}x/\ell_p} R(\theta)  + \sum_{k>0}a_{k}e^{\lambda _{k}x/\ell_p}\Theta_k+\sum_{k<0}a_{k}e^{\lambda _{k}\left( x-L\right)/\ell_p}\Theta_k.
\label{eq:6-two-way-sol-expansion-force}
\end{equation}

In analogy with the zero force case, the functions $R(\theta)$, $1$, and $\{\Theta_k\}_{k=-\infty}^{\infty}$ are complete on $(-\pi, \pi)$ \cite{Beals1981} and obey orthogonality relations
\begin{align}
 &\int_{-\pi}^{\pi} \Theta_k (\cos \theta - r) d\theta = 0 \\
 &\int_{-\pi}^{\pi} \Theta_k R(\theta) (\cos \theta - r) \theta d\theta = 0  \hspace{8mm} \\
 &\int_{-\pi}^{\pi} \Theta_j \Theta_k (\cos \theta - r) d\theta = \text{sgn}(j) \delta_{jk}, \hspace{8mm} (j,k) \neq (0,0)
\end{align}
assuming appropriate normalization. Using these relations, the iterative procedure from Section III can be applied, provided at each iteration the diffusion solution is replaced by $\beta e^{\lambda_{R}x/\ell_p} R(\theta)$ (this ensures that the zero force case is smoothly recovered as $r \rightarrow 0$).

\textbf{Sedimentation profile}

To study the behavior of sedimenting ABPs, we consider particles on a semi-infinite interval $0 < x < \infty$ and with boundary data $f(0, \theta) = v_+(\theta)$. For the boundary condition at $x = \infty$ we assume that $f(x, \theta) \rightarrow 0$ as $x \rightarrow \infty$, which agrees with simulations over accessible timescales. Enforcing the latter constraint, Eq. (\ref{eq:6-two-way-sol-expansion-force}) becomes
\begin{equation}
f(x, \theta) =  \beta e^{\lambda_{R}x/\ell_p} R(\theta)  + \sum_{k>0}a_{k}e^{\lambda _{k}x/\ell_p}\Theta_k;
\label{two-way-sol-expansion-semi-infinite}
\end{equation}
the remaining coefficients are determined in order to satisfy $f(0, \theta) = v_+(\theta)$ where $\cos \theta - r > 0$. To solve the problem, therefore, we must be able to express a general function $v_+(\theta)$ on the partial range $\cos \theta - r > 0$ using only \emph{half} of the spectrum.

\emph{Distal profile.}
Previous theoretical work on the sedimentation of active particles has demonstrated an exponential decay of the density $g(x) = \int f(x , \theta) d\theta$ for distances far from $x = 0$ \cite{Tailleur2009, Enculescu2011,Solon2015c}, a trend also observed experimentally \cite{Palacci2010}. This behavior is easily recovered in the framework here: The boundary layer contributions inside the sum in Eq. (\ref{two-way-sol-expansion-semi-infinite}) are negligible except near $x = 0$, leaving $g(x) \propto  e^{-\lambda_{R}x/\ell_p}$. Incidentally, we note that while the appearance of an exponential decay might suggest a Boltzmann form $e^{-V(x)/T_{eff}}$ with an effective temperature $T_{eff}$, this is only valid if the cubic corrections to $\lambda_R$ are negligible, as already discussed by Solon, et al \cite{Solon2015c}.

\emph{Proximal profile.} Near the wall, the terms inside the sum in Eq. (\ref{two-way-sol-expansion-semi-infinite}) cannot be neglected. We note, however, that although the half spectrum in Eq. (\ref{two-way-sol-expansion-semi-infinite}) is expected to be complete where $\cos \theta - r > 0$ (see Ref. \cite{Beals1981} for examples of such \emph{half-range expansion} theorems), the eigenfunctions are not orthogonal on this interval, thereby precluding straightforward determination of the coefficients. Instead, an iterative procedure must be employed. While the iterative procedure from Section III as constructed applies only on a finite interval, we can nevertheless calculate the coefficients in Eq. (\ref{two-way-sol-expansion-semi-infinite}) to arbitrary precision by first considering the problem on $0 < x < L$ with boundary data
\begin{eqnarray}
f(0,\theta ) &=&v_+(\theta),\text{ \ \ \ \ \ where } \cos(\theta) - r >0 \\
f(L,\theta ) &=&0,\text{ \ \ \ \ \ \ \ \ \ \ \ where } \cos(\theta) - r <0
\end{eqnarray}
and calculating the corresponding expansion coefficients in Eq. (\ref{eq:6-two-way-sol-expansion-force}) using the ordinary iterative procedure. If this procedure is executed for arbitrarily large $L$, we expect to recover the semi-infinite expansion given by Eq. (\ref{two-way-sol-expansion-semi-infinite}), a conclusion which is supported by explicit numerical calculation.

\begin{figure}
  \includegraphics[width=0.4347826087\linewidth,height=.326\linewidth]{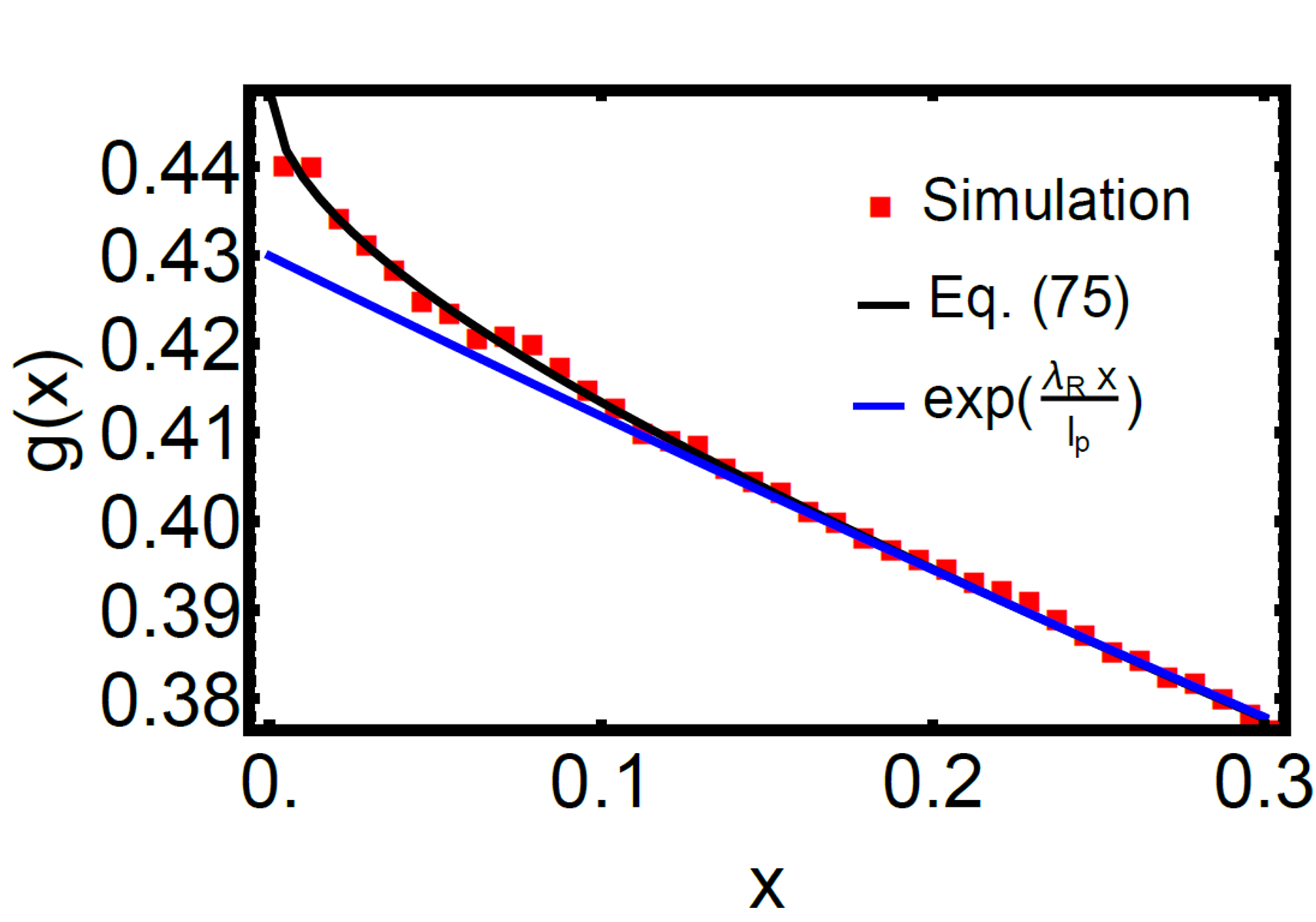}
  \includegraphics[width=0.4347826087\linewidth,height=.3\linewidth]{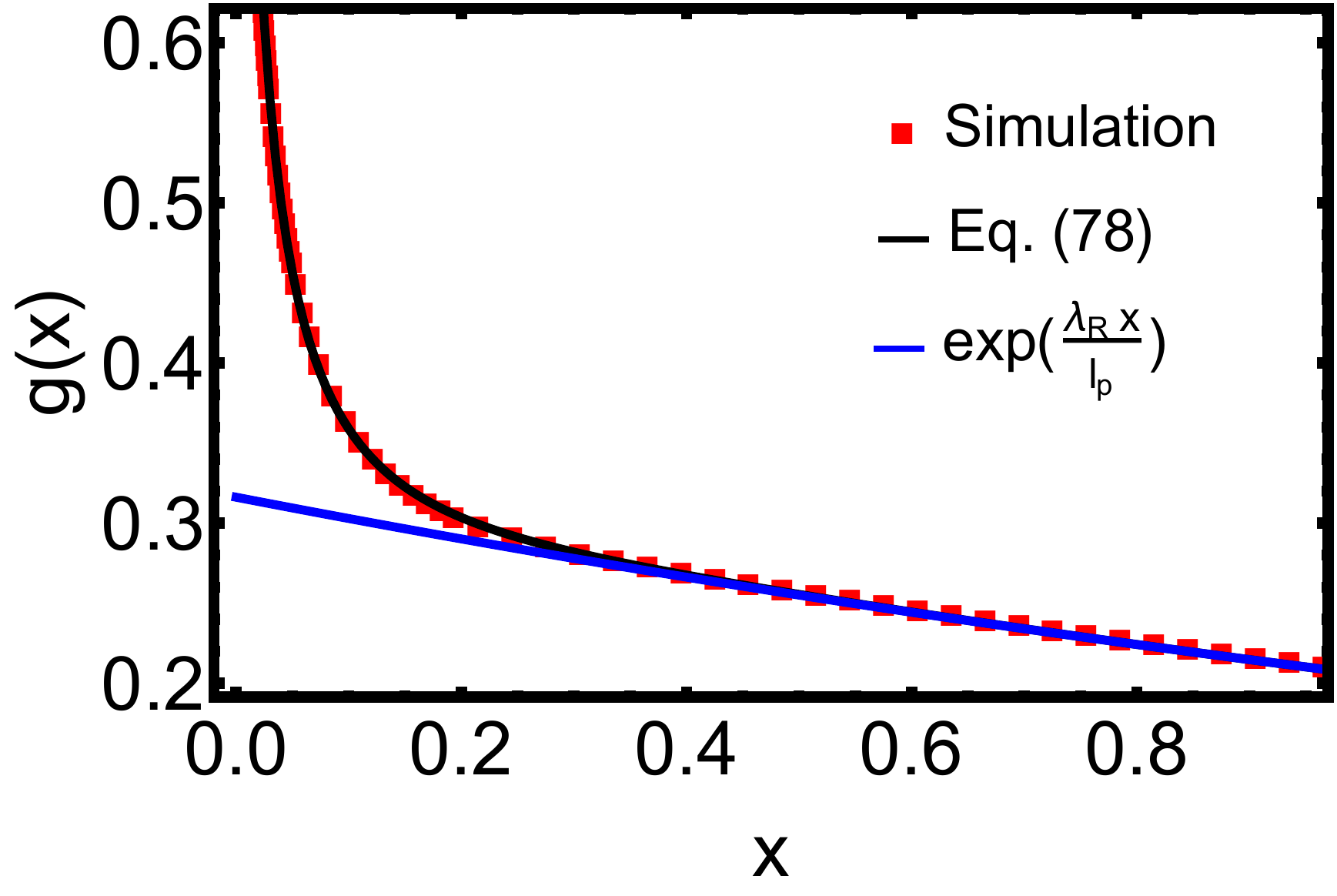}
  \caption{Density profile of sedimenting ABPs assuming a uniform reservoir of ABPs at $x=0$ (left) or a hard wall (right), and in both cases taking $r= 0.2$. The density calculated using the iterative procedure (black) is compared with the distal part (blue), thereby identifying a proximal region near $x = 0$ where the surface layer terms in Eq. (\ref{two-way-sol-expansion-semi-infinite}) are non-negligible. The density is calculated on the left from the iterative procedure using 100 eigenfunctions and 10 iterations, and on the right by fitting Eq. (\ref{eq:hard-wall-proximal-profile}). The values of the fit parameters in the latter case are $b \simeq 2.92$ and $c \simeq 0.316$. The overall normalization in both cases is arbitrary. }
  \label{fig:proximal_dist}
\end{figure}

The generic effect of the boundary appears to be a local increase of the density near $x = 0$. For instance, the results of choosing a uniform reservoir of ABPs at $x = 0$ (such that $v_+(\theta) = \rho_1$, independent $\theta$) are shown in Fig. \ref{fig:proximal_dist}, demonstrating a distinct increase of density starting near $x = 0.1 \ell_p$. More dramatic is the profile near a hard wall boundary of the same type used in the infinite channel calculation (c.f. Fig. \ref{fig:infinite_channel}), in which case the near-wall corrections manifest as a power law divergence. Although the iterative procedure appears to converge relatively slowly in this case, we can nevertheless study the form of the divergence using the approximate scaling
\begin{equation}
g(x) \sim \frac{a}{\sqrt{x}} \left[1 -  \text{erf}\left(b\sqrt{\frac{x}{\ell_p}}\right)\right] + c  e^{\lambda_R x/\ell_p},
\label{eq:hard-wall-proximal-profile}
\end{equation}
which can be derived from Eq. \ref{two-way-sol-expansion-semi-infinite} (see Appendix B). Here $a$, $b$, and $c$ are constants to be determined by fitting or using the iterative procedure. Fitting to simulation data (Fig. \ref{fig:proximal_dist}) shows that this form is quite accurate for all values of $x$. Although the strict divergence is regularized in realistic systems due to translational diffusion, noticeable particle accumulation near the wall is expected to persist \cite{Elgeti2013} and is likely to be experimentally observable.

\section{Application to other active particle models}

The techniques developed here are not limited to ABPs. More generally, we may consider an active particle model for which the stationary distribution is given by an equation of the form
\begin{equation}
h(\theta) \frac{\partial f(x,\theta)}{\partial x} = \hat{R}_{\theta} f(x,\theta)
\label{general-two-way}
\end{equation}
where $\hat{R}_{\theta}$ is an operator describing particle reorientations, and $h(\theta)$ quantifies the particles' self-propulsion. Periodic boundary conditions are assumed in $\theta$. If we can find a function $g(\theta)$ satisfying the periodicity boundary condition and
\begin{equation}
\hat{R}_{\theta} g(\theta) = h(\theta),
\label{eq:general_diff_sol}
\end{equation}
then it is easy to verify that Eq. (\ref{general-two-way}) possesses the diffusion solution $\beta(x + g(\theta))$. With appropriate mild constraints on $\hat{R}_{\theta}$, it then makes sense to expand the general solution as
\begin{equation}
f(x,\theta) = \alpha + \beta(x + g(\theta)) + \sum_{k>0}a_{k}e^{\lambda_{k}x}\phi_k(\theta)+\sum_{k<0}a_{k}e^{\lambda_{k}\left( x-L\right)} \phi_k(\theta)
\label{general-two-way-expansion}
\end{equation}
with eigenfunctions and eigenvalues given by $\hat{R}_{\theta} \phi_k(\theta) = \lambda_k h(\theta) \phi_k(\theta)$. The expansion coefficients can be calculated using the iterative procedure developed in Section III, leading to steady-state distributions very similar in form to those studied for ABPs. Thus, the behaviors derived for ABPs are expected to extend to a variety of active particle models.

As we have already observed in the case of sedimenting ABPs, however, not all systems give rise to a diffusion solution. This is more generally seen in the case of particles with biased self-propulsion, such that $\int_{-\pi}^{\pi} h(\theta) d\theta \neq 0$ (and taking $\hat{R}_{\theta} = \partial^2/\partial \theta^2$ for definiteness). In this model, the diffusion solution no longer exists because Eq. (\ref{eq:general_diff_sol}) does not have a solution satisfying boundary conditions. On the other hand, this fact does not invalidate the expansion (\ref{general-two-way-expansion}); Beals \cite{Beals1981,Beals1981E} has shown that even under these general circumstances the remaining terms in the expansion span the solution space by themselves.

We close by discussing the applicability of the techniques developed here to AOUPs, particles whose activity is modeled as Gaussian colored noise. This model has received much recent attention in part due to its solvability through various approximation schemes, leading to important insights about the collective behavior of active systems \cite{Maggi2015,Marconi2015}. Under confinement, the techniques developed here may be used to find a formally exact solution.

In one dimension, the steady-state Smoluchowski description for AOUPs can be written as
\begin{equation}
w \frac{\partial f(x,w)}{\partial x} = \frac{\partial}{\partial w}\left[ w f(x,w) \right] + \ell_p^2 \frac{\partial^2 f(x,w)}{\partial w^2}
\label{GCN_FP}
\end{equation}
where $w$ is a degree of freedom parametrizing the direction and magnitude of the self-propulsion, and $\ell_p$ is an effective persistence length. We remark that this equation has the same structure as the steady-state Fokker-Planck equation for a Brownian particle in phase space, with $w$ playing the role of velocity. In that context the solution in the presence of boundaries can provide insight into kinetic boundary layer effects of diffusion processes. With this goal, previous work has sought to solve the problem on the semi-infinite domain by treating it as a two-way diffusion problem (e.g. \cite{Titulaer1984,Burschka1981,Selinger1984}). The problem on the finite domain can be treated using the iterative technique developed here. We briefly outline the steps. First, making the transformation $f(x,w) = \text{exp}\left[-\frac{1}{2} \left( \frac{w}{\ell_p} \right)^2\right] g(x,w)$, Eq. (\ref{GCN_FP}) becomes
\begin{equation}
w \frac{\partial g(x,w)}{\partial x} = \ell_p^2 \frac{\partial^2 g(x,w)}{\partial w^2} - w \frac{\partial g(x,w)}{\partial w}.
\end{equation}
We again write the general solution as a sum of separable solutions plus a diffusion solution. The necessary spectral theorem regarding the completeness of the functions in this expansion is provided by Beals and Protopopescu, who also give an explicit form for the separable solutions \cite{Beals1983}. We therefore obtain the form of the general solution as (in units where $\ell_p = 1$)
\begin{equation}
g(x,w) = \alpha + \beta(w - x) + \sum_{k>0}a_{k}e^{-\lambda_{k}x}u_k(w)+\sum_{k<0}a_{k}e^{-\lambda_{k}\left( x-L\right)} u_k(w)
\label{GCN_sol_expansion}
\end{equation}
with eigenfunctions and eigenvalues given by
\begin{align}
&\lambda_{\pm k} = \pm \sqrt{k}, \hspace{8mm} k = 1, 2, 3, \ldots  \\
&u_{\pm k}(w) = C_k e^{\pm w \sqrt{k}} H_k\left(\frac{w}{\sqrt{2}} \mp \sqrt{2k} \right), \hspace{8mm} k = 1, 2, 3, \ldots \\
&C_k = \left[ (8 \pi k)^{1/4} 2^{k/2} e^{k} \sqrt{k!}  \right]^{-1}
\end{align}
and $H_k(z)$ the $k^{th}$ Hermite polynomial. The $u_{\pm k}$ satisfy the orthogonality relations
\begin{align}
&\int_{-\infty}^{\infty} w u_{j} u_{k}  e^{-w^2/2} dw = \text{sgn}(j) \delta_{jk}  \hspace{8mm} \\
&\int_{-\infty}^{\infty} w u_{k} e^{-w^2/2} dw = \int_{-\infty}^{\infty} w^2 u_{k} e^{-w^2/2} dw = 0.
\end{align}
We note that far from the boundaries and in the absence of any net flux, the distribution of the velocity variable $w$ takes on a Gaussian form, analogous to the equilibrium velocity distribution. Given proper boundary conditions (i.e., which specify $f(0, w)$ for $w \in (0, \infty)$ and $f(L, w)$ for $w \in (-\infty, 0)$), the constants in Eq. (\ref{GCN_sol_expansion}) can be determined using the iterative procedure from Section III. Because the $u_k$ with $k > 0$ are exponentially small on the negative range ($w < 0$) in comparison with the positive range ($w > 0$), and vice versa for the eigenfunctions with $k < 0$, the iterative procedure is expected to converge by the same argument from Section III.

\section{Summary}

We have developed a formally exact technique for obtaining the steady-state distributions of ABPs in a range of physical scenarios. The resulting distributions can be readily analyzed to explain a variety of behaviors. In particular, we have precisely quantified the intriguing effects of confinement on ABPs, including accumulation at walls and orientational order in the absence of aligning interactions. Moreover, by considering a constant flux steady state, we have obtained an effective diffusivity that carries signatures of the persistent motion characterizing active particle trajectories. Finally, we have calculated the proximal part of the density profile for sedimenting ABPs, and have demonstrated the appearance of a power-law divergence for the case of a hard wall boundary. The techniques developed are applicable to a wide range of models for active particles.





\section*{Acknowledgments}
We thank R. Beals for helpful discussions regarding various aspects of two-way diffusion equations. This work was supported by the NSF (DMR-1149266, the Brandeis MRSEC DMR-1420382, and IGERT DGE-1068620).
Computational resources were provided by the NSF through XSEDE computing resources and the
Brandeis HPCC.

\section*{Appendix A: Construction of the angular eigenfunctions}

We detail the construction of the odd eigenfunctions in the zero force case; the remaining eigenfunctions (including those for a nonzero force) can be constructed following the same procedure. Our first step is to introduce the $2\pi$ periodic Fourier series:
\begin{equation}
\Theta(\theta) = \sum_{n=1}^{\infty} a_n \sin(n \theta),
\label{eq:A-fourier-ansatz}
\end{equation}
and substitute into
\begin{equation}
\frac{d^{2}\Theta}{d\theta ^{2}}=\lambda (\cos \theta) \Theta.
\label{eq:A-theta-eigenvalue-eqtn}
\end{equation}
The Fourier Ansatz satisfies the periodicity boundary conditions at the outset, thus giving the appearance that $\lambda$ is unconstrained. This is incorrect, however, because we cannot generally assume the convergence of the series in Eq. (\ref{eq:A-fourier-ansatz}); in fact, it converges only for special values of $\lambda$, which is precisely the spectrum we seek. Let us therefore identify these values of $\lambda$ and solve for the corresponding Fourier coefficients. Substituting into Eq. (\ref{eq:A-theta-eigenvalue-eqtn}), multiplying by $\sin(m \alpha)$, and integrating gives rise to the recurrence relation
\begin{align}
a_{m+1}&=-\frac{2m^{2}}{\lambda }a_{m}-a_{m-1}, \; \; \; \; \; \; m>1  \label{eq:A-recursion1} \\
 a_{2}&=-\frac{2}{\lambda }a_{1}.
\label{eq:A-recursion0}
\end{align}
Experimenting with this recurrence using arbitrarily chosen values of $\lambda$ and $a_1$, one finds that the terms generally diverge rapidly. The convergent solutions take more work to find. We may argue generally as follows: Since Eq. (\ref{eq:A-recursion1}) constitutes a second order recurrence, the general solution may be written as a sum of any two linearly independent solutions $f_m$ and $g_m$:
\begin{equation}
a_m = A f_m + B g_m
\end{equation}
with $A$ and $B$ determined self-consistently from $a_1$ and $\lambda$ and application of Eqs. (\ref{eq:A-recursion1}) and (\ref{eq:A-recursion0}). In particular, for the given recurrence it is possible to prove (for example, using Pincherle's theorem \cite{Gautschi1967}) that the decomposition may be chosen such that $\lim_{m\rightarrow \infty }f_{m} / g_{m} = 0$, i.e. the recurrence possesses a subdominant solution which we call $f_m$. Therefore, $f_m$ is our candidate for a convergent solution, and is obtained by self-consistently selecting $\lambda$ such that $B=0$.

However, the particular solution $f_m$ is inaccessible by direct numerical iteration. This is because even if we happen to know to (say) ten digits the value of $\lambda$ for which $B = 0$, the error in the 11th digit will still contaminate the solution with a piece of $g_m$,
\begin{equation}
a_m = A f_m + \epsilon g_m,
\end{equation}
and because $\lim_{m\rightarrow \infty }f_{m} / g_{m} = 0$, no matter how small $\epsilon$, the solution will eventually be dominated by $g_m$, which is divergent. Therefore, this numerical algorithm will not work, besides not telling us in a simple manner what $\lambda$ is.

A better approach utilizes the close relationship between second order recurrences and the eigenvectors of a tridiagonal matrix. We see that Eqs. (\ref{eq:A-recursion1}) and (\ref{eq:A-recursion0}) can be written in matrix form as
\begin{equation}
\left(
\begin{array}{cccccc}
0 & -1/2 & 0 & 0 & 0 & \cdots  \\
-1/8 & 0 & -1/8 & 0 & 0 & \cdots  \\
0 & -1/18 & 0 & -1/18 & 0 & \cdots  \\
0 & 0 & -1/32 & 0 & -1/32 & \cdots  \\
0 & 0 & 0 & -1/50 & 0 & \ddots  \\
\vdots  & \vdots  & \vdots  & \vdots  & \ddots  & \ddots
\end{array}%
\right) \left(
\begin{array}{c}
a_{1} \\
a_{2} \\
a_{3} \\
a_{4} \\
a_{5} \\
\vdots
\end{array}%
\right) =\frac{1}{\lambda }\left(
\begin{array}{c}
a_{1} \\
a_{2} \\
a_{3} \\
a_{4} \\
a_{5} \\
\vdots
\end{array}%
\right).
\end{equation}

Now suppose we truncate this system at finite dimension $n \times n$,  and solve for the eigenvectors and eigenvalues. We see that the eigenvalues obtained are precisely the values of $\lambda$ that give rise to a solution of Eq. (\ref{eq:A-recursion1}) defined on $m = 1, 2, \ldots, n+1$ and satisfying the boundary condition $a_{n+1} = 0$. It follows that taking $n \rightarrow \infty$, the eigenvalues are the values of $\lambda$ for which the solution to Eq. (\ref{eq:A-recursion1}) satisfies $\lim_{n\rightarrow \infty }a_{n}=0$; the eigenvectors then correspond precisely to the desired subdominant solution. In practice, efficient numerical algorithms for computing the spectrum of a tridiagonal matrix are available in computer algebra systems.

Finally, we briefly review the normalization scheme used in the main text. First, we normalize the $\Theta_k$ for $k > 0$ such that
\begin{equation}
\int_{-\pi}^{\pi} \Theta_k^2 \cos \theta d\theta = 1,  \hspace{8mm} k > 0.
\label{eq:A-norm1}
\end{equation}
For $k < 0$, Eq. (\ref{eq:A-theta-eigenvalue-eqtn}) implies that we can take $\Theta_{k}(\theta + \pi) = \Theta_{-k}(\theta)$. Translating the integral in Eq. (\ref{eq:A-norm1}) by $\pi$ and applying this relation gives
\begin{equation}
\int_{-\pi}^{\pi} \Theta_k^2 \cos \theta d\theta = -1,  \hspace{8mm} k < 0,
\label{eq:A-norm2}
\end{equation}
which recovers the normalization in the main text.


\section*{Appendix B: Scaling of the excess density}

We consider first the infinite channel in the zero force case, and work in terms of the excess density $g_{E}(x) = \int f(x,\theta) d\theta - 2 \pi A \alpha$. Expanded out, this is
\begin{equation}
g_{E}(x) = \sum_{k>0}A_{k}e^{\lambda _{k}x}+\sum_{k<0}A_{-k}e^{\lambda _{k}\left( x-L\right) }
\label{eq:B-reduced-excess}
\end{equation}
where $A_k = a_k \int_{-\pi}^{\pi} \Theta_k d\theta$. Focusing on the left wall, we assume $L$ is large enough such that the second sum is negligible. Moreover, because of the symmetry of the boundary conditions, only the even eigenfunctions appear in the above sum. Thus, we only need to consider the eigenfunctions having negative eigenvalues and even parity. Calculating the $A_k$ using the iterative procedure shows that they are very nearly equal to some constant $A$. In fact, this is expected: In analogy with an ordinary Fourier series, the expansion of a delta function as a series in $\Theta_k$ produces coefficients of $\mathcal{O}(1)$. We now examine what this observation implies about the scaling of $g_{E}(x)$.

To approximate the scaling of $g_{E}(x)$, we replace the sum over $k$ with an integral:
\begin{equation}
g_{E}(x) \sim A\int_{1}^{\infty}e^{\lambda _{k}x} dk.
\label{eq:B-integral-approx}
\end{equation}
This integral can be evaluated if we know how the eigenvalues scale with $k$. In fact, using a WKB approximation, it is not difficult to show that (in units where $\ell_p = 1$) the negative eigenvalues of the even eigenfunctions are given approximately by
\begin{align}
\int_{-\pi / 2}^{\pi / 2} \sqrt{\lambda_k \cos \theta} d\theta = -\left(2k + \frac{1}{2} \right) \pi \\
\rightarrow \lambda_k \simeq -0.4297 (4k + 1)^2
\label{eq:B-WKB-scaling}
\end{align}
where the expression is asymptotic in $k$. Substituting into Eq. (\ref{eq:B-integral-approx}) and evaluating the integral gives
\begin{equation}
g_E(x) \sim \frac{a}{\sqrt{x}}\left[1 - \text{erf}(3.28\sqrt{x})\right]
\label{eq:B-reduced-density-power-law0}
\end{equation}
where $a$ is a constant. This result does not exactly match the trend observed from the exact expression for $g_E(x)$ in Eq. (\ref{eq:B-reduced-excess}), most likely because of the error in Eqs. (\ref{eq:B-integral-approx}) and (\ref{eq:B-WKB-scaling}) for small $k$. A slightly better fit (found by eye) is given by
\begin{equation}
g_E(x) \sim \frac{a}{\sqrt{x}}\left[1 - \text{erf}(2.17\sqrt{x})\right].
\label{eq:B-reduced-density-power-law1}
\end{equation}
The same procedure can be applied to the expansion in the constant force case, recovering Eq. (\ref{eq:hard-wall-proximal-profile}). The crucial similarity in this case is that the eigenvalues still scale quadratically with $k$, thus resulting in a form similar to Eq. (\ref{eq:B-reduced-density-power-law1}).

\section*{Appendix C: Simulation details}
\label{sec:appendixC}
Simulation results were obtained by directly sampling particle trajectories from the stochastic dynamics given by Eqs. (\ref{eq:2-abp-spatial}) and (\ref{eq:2-abp-rotational}) (Eqs. (\ref{eq:6-abp-spatial-force}) and (\ref{eq:6-abp-rotational-force}) in the case of sedimentation). In units where $D_r = v_0 = 1$, the time step $\Delta t$ was chosen for most simulations to be $0.0001$. For narrow channels ($L < 0.1$), a smaller timestep was chosen. The simulations were run using at least $60000$ particles initialized with random positions and orientations. To be certain that steady state was achieved, the simulations were run for roughly $10^7$ timesteps depending on channel width (or depending on the sedimentation length in the case of a constant force), which would be sufficient to guarantee the achievement of steady state even if the particles lacked self-propulsion. Data was subsequently collected for another $\sim 10^7$ timesteps.

\bibliographystyle{apsrev4-1}
\bibliography{bib}

\end{document}